\documentclass{aa}  
\usepackage{graphicx}
\usepackage{txfonts}
\begin{document}

   \title{GMASS ultradeep spectroscopy of galaxies at $z \sim 2$. IV. 
          The variety of dust populations}
   \author{S. Noll\inst{1,2}
   \and    D. Pierini\inst{2}
   \and    A. Cimatti\inst{3}
   \and    E. Daddi\inst{4}
   \and    J. D. Kurk\inst{5}
   \and    M. Bolzonella\inst{6}
   \and    P. Cassata\inst{7}
   \and    C. Halliday\inst{8}
   \and    M. Mignoli\inst{6}
   \and    L. Pozzetti\inst{6}
   \and    A. Renzini\inst{9}
   \and    S. Berta\inst{2}
   \and    M. Dickinson\inst{10}
   \and    A. Franceschini\inst{11}
   \and    G. Rodighiero\inst{11}
   \and    P. Rosati\inst{12}
   \and    G. Zamorani\inst{6}}

   \offprints{S. Noll}

   \institute{Observatoire Astronomique de Marseille-Provence, 38 rue 
              Fr\'ed\'eric Joliot-Curie, 13388 Marseille Cedex 13, France\\
              \email{stefan.noll@oamp.fr}
   \and       Max-Planck-Institut f\"ur extraterrestrische Physik, 
              Giessenbachstr., 85748 Garching, Germany
   \and       Dipartimento di Astronomia, Universit\`a di Bologna, 
              Via Ranzani 1, 40127 Bologna, Italy
   \and       CEA-Saclay, DSM/DAPNIA/Service d'Astrophysique, 
              91191 Gif-sur-Yvette Cedex, France; 
              AIM-Unit\'e Mixte de Recherche CEA-CNRS (\#7158)-Universit\'e 
              Paris VII, France
   \and       Max-Planck-Institut f\"ur Astronomie, K\"onigstuhl 17, 
              69117 Heidelberg, Germany              
   \and       INAF - Osservatorio Astronomico di Bologna, Via Ranzani 1, 
              40127 Bologna, Italy
   \and       Department of Astronomy, University of Massachusetts, LGRT-B 
              619E, 710 North Pleasant Street, Amherst, MA 01003-9305, USA
   \and       INAF - Osservatorio Astrofisico di Arcetri, Largo E. Fermi 5,
              50125 Firenze, Italy
   \and       INAF - Osservatorio Astronomico di Padova, Vicolo
              dell'Osservatorio 5, 35122 Padova, Italy
   \and       NOAO-Tucson, 950 North Cherry Avenue, Tucson, AZ 85719, USA
   \and       Universit\`a di Padova, Dipartimento di Astronomia, Vicolo
              dell'Osservatorio 2, 35122 Padova, Italy
   \and       European Southern Observatory, Karl-Schwarzschild-Str. 2, 
              85748 Garching, Germany 
}

   \date{Received / Accepted}

 
  \abstract
   {}
{The properties of dust attenuation at rest-frame UV wavelengths are inferred 
from very high-quality FORS\,2 spectra of 78 galaxies from the GMASS survey 
at redshift $1 < z < 2.5$. These objects complement a previously investigated 
sample of 108 UV-luminous, intermediate-mass (i.e., with stellar masses 
$\sim 10^{10} - 10^{11}\,\mathrm{M}_{\sun}$) galaxies at similar redshifts, 
selected from the FDF spectroscopic survey, the K20 survey, and the GDDS. 
Detection of the broad absorption feature centred on about 2175\,\AA{} 
(``UV bump'') implies that the average UV extinction curve of a galaxy more 
closely resembles that of the Milky Way (MW) or the Large Magellanic Cloud 
(LMC), and differs from that of the Small Magellanic Cloud (SMC).}
{The shape of the UV extinction curve is constrained by a parametric 
description of the rest-frame UV continuum with the support of a suite of 
models combining radiative transfer and stellar population synthesis. The UV 
bump is further characterised by fitting Lorentzian-like profiles.}
{Spectra exhibit a significant 2175\,\AA{} feature in at least 30\% of the 
cases, especially those suffering from substantial reddening. If attenuation 
is dominated by dust ejected from the galaxy main body via galactic winds or 
more localised superwinds, UV extinction curves in-between those of the SMC 
and LMC characterise UV-luminous, intermediate-mass galaxies at 
$1 < z < 2.5$. The fraction of galaxies with extinction curves differing from 
the SMC one increases, if more dust resides in the galactic plane or dust 
attenuation depends on stellar age. On average, the width of the manifested 
UV bumps is about 60\% of the values typically measured along sightlines in 
the LMC and MW. This suggests the presence of dust similar to that found in 
the ``supergiant'' shell of ionised filaments LMC\,2, close to 30\,Dor. The 
presence of the carriers of the UV bump (probably organic carbon and 
amorphous silicates) at $1 < z < 2.5$ argues for outflows from asymptotic 
giant branch (AGB) stars being copious then. Consistent with their higher 
star-formation rates, the GMASS galaxies with a manifested UV bump are more
luminous at rest-frame 8\,$\mu$m, where the emission is dominated by
polycyclic aromatic hydrocarbons (also products of AGB stars). In addition,
they exhibit larger equivalent widths for prominent UV (metal) absorption
features, mostly of interstellar origin, which indicates overall more evolved
stellar populations.}
{We conclude that diversification of the small-size dust component
(responsible for the differential extinction at UV wavelengths and the
emission at mid-IR wavelengths) has already started in the most evolved
star-forming systems at $1 < z < 2.5$.}


   \maketitle
%

\section{Introduction}\label{introduction}

A broad absorption excess centred on 2175\,\AA{} (the so-called ``UV bump'')
characterises the extinction curves usually inferred for sightlines towards 
the Milky Way (MW) and Large Magellanic Cloud (LMC) (for reviews see 
Fitzpatrick \cite{FIT04} and Clayton \cite{CLA04}), and also towards M\,31 
(Bianchi et al. \cite{BIA96}). In contrast, this feature is absent in the 
typical extinction curves of the Small Magellanic Cloud (SMC), which exhibit
a steep rise at far-UV wavelengths. SMC-like dust appears to characterise
starburst galaxies in the local Universe (Calzetti et al. \cite{CAL94})
and Lyman break galaxies (LBGs) at redshifts $2 < z < 4$ (Vijh et al. 
\cite{VIJ03}). For this reason, the so-called ``Calzetti law'' (Calzetti et 
al. \cite{CAL94}, \cite{CAL00}) is frequently assumed to describe attenuation 
by dust in starburst and star-forming galaxies in the distant Universe.

However, there is strong support for the presence of the 2175\,\AA{} feature
in spectra of at least 30\% of the vigorously star-forming galaxies at 
$2 < z < 2.5$ (Noll \& Pierini 2005, hereafter referred to as NP05) and 
$1 < z < 1.5$ (Noll et al. 2007, hereafter referred to as NPPS). Furthermore, 
an LMC-like extinction curve with a UV bump was inferred clearly from both 
the spectrum and the broad-band spectral energy distribution (SED) of the 
afterglow of the gamma ray burst GRB\,070802 at $z = 2.45$ (Fynbo et al. 
\cite{FYN07}; Kr\"uhler et al. \cite{KRU08}; El\'iasd\'ottir et al. 
\cite{ELI08}). As for other objects at intermediate/high redshifts, the 
presence of the carriers of the 2175\,\AA{} feature is controversial in 
quasars/active galactic nuclei (AGN) (cf. Pitman et al. \cite{PIT00}; 
Maiolino et al. \cite{MAI01}; Hopkins et al. \cite{HOP04}; Vernet et al. 
\cite{VER01}) and intervening Mg\,II absorbers (Malhotra \cite{MAL97}; Wang 
et al. \cite{WAN04}; Vanden Berk et al. \cite{VAN08}), whereas it is likely 
in Ca\,II absorbers at intermediate redshifts (Wild \& Hewett \cite{WIL05}).

It would therefore be important and interesting to understand when the 
carriers of the UV bump first appeared in the Universe and how they can 
survive in vigorously star-forming galaxies at high redshifts. Multiple 
carriers (i.e., organic carbon and amorphous silicates) are likely to explain 
the almost invariant central wavelength and variable bandwidth of the 
2175\,\AA{} feature (Bradley et al. \cite{BRA05} and references therein). 
However, a random, hydrogen-free assembly of microscopic sp$^2$ carbon chips, 
forming a macroscopically homogeneous and isotropic solid, has recently been 
proposed as a model carrier of the UV bump by Papoular \& Papoular 
(\cite{PAP09}). According to this promising model, different grain formation 
histories and/or different interstellar radiation fields and other heating 
effects are responsible for a change in bandwidth without notably impacting 
the central wavelength. Whatever the carrier of the UV bump, the proportions 
of the interstellar dust ingredients that are responsible for both the UV 
bump and the far-UV rise in the extinction curve, relative to each other, are 
approximately similar in both the MW and LMC, but appear to differ 
significantly in the SMC (Whittet \cite{WHI03}). Differences also exist 
within the MW for different sightlines and towards different environments, 
from dense molecular clouds to the diffuse interstellar medium (ISM; see 
Fitzpatrick \cite{FIT04}). They must reflect the extreme sensitivity of the 
small-size dust components to the local chemical enrichment and energy budget 
(in terms of radiation field and shocks), as well as their selective removal 
from the size distribution due to a number of physical processes (e.g., 
Whittet \cite{WHI03}; Gordon et al. \cite{GOR03}).

We continue an analysis started by \cite{NOL05} and extended by \cite{NOL07}. 
We adopt a new homogeneous sample of UV-luminous, intermediate-mass galaxies 
at $1 < z < 2.5$ drawn from the spectroscopic catalogue of the galaxy mass 
assembly ultra-deep spectroscopic survey (GMASS, Kurk et al. \cite{KUR08A}; 
Cimatti et al. \cite{CIM08}; Kurk et al., in prep.) in the GOODS-South Field 
(GOODS-S; Giavalisco et al. \cite{GIA04}). This sample has about 70\% of the 
size of the entire sample in \cite{NOL07}, but contains mostly objects in the 
poorly investigated range $1.5 < z < 2.0$, known as the ``redshift desert''. 
The excellent dataset available for the GMASS sample enables us to 
characterise more accurately the apparent shape of the 2175\,\AA{} feature in 
high-redshift star-forming galaxies, and to investigate the relation between 
dust properties inferred from attenuation at UV wavelengths and emission at 
mid-IR wavelengths. Furthermore, it adds relevant information about the link 
between small-size dust populations (i.e., large molecules and small grains) 
and chemical abundances in stars and gas of the host galaxies.

Throughout this paper, we adopt a $\Lambda$CDM cosmological model where 
$\Omega_{\Lambda} = 0.7$, $\Omega_\mathrm{M} = 0.3$, and 
$H_0 = 70\,\mathrm{km}\,\mathrm{s}^{-1}\,\mathrm{Mpc}^{-1}$.
By default, photometry is given in the Vega magnitude system; otherwise AB 
magnitudes are indicated.

\section{The sample}\label{sample}

\subsection{The GMASS sample}\label{gmass}

The GMASS spectroscopic catalogue (Kurk et al. \cite{KUR08A}; Kurk et al., in 
prep.) comprises 167 objects with secure spectroscopic redshifts. It is based 
on a mass-sensitive selection criterion using photometry in the {\em Spitzer} 
$4.5$\,$\mu$m IRAC filter ($m_\mathrm{AB}(4.5\,\mu{\rm m}) < 23$), and the 
additional photometric-redshift criterion $z_\mathrm{phot} > 1.4$. For our 
analysis, we identify and select vigorously star-forming objects, for which 
we can robustly describe the continuum emission at rest-frame wavelengths 
from 1700 to 2600\,\AA{} in the parametric way illustrated by \cite{NOL05} 
and \cite{NOL07} (see Sect.~\ref{parameters}). In the (spectroscopic) 
redshift range of our interest ($1 < z < 2.5$), GMASS yields a total of 78 
class~2 objects, i.e., actively star-forming galaxies with UV spectra
dominated by young stellar populations and no AGN features (see Mignoli et 
al. \cite{MIG05}). For 68 objects spectroscopically confirmed to lie at 
$1.5 < z < 2.5$, selection biases are understood: the average galaxy is about 
$0.2$--$0.25$\,mag fainter at $4.5~\mu$m 
($\langle m_\mathrm{AB}(4.5\,\mu \mathrm{m}) \rangle = 22.0$, $\sigma = 0.7$) 
than the average objects in the full spectroscopic and photometric samples 
for $m_\mathrm{AB}(4.5\,\mu \mathrm{m}) < 23$ and $1.5 < z < 2.5$. 
Conversely, it is about $0.3$ and $0.8$\,mag brighter in the visual 
($\langle V_{606,\mathrm{AB}} \rangle = 24.3$, $\sigma = 0.5$) than the 
average galaxies in these two catalogues, respectively. Therefore, the 
following analysis addresses almost exclusively the sample of 68 GMASS 
UV-luminous galaxies at $1.5 < z < 2.5$.

The spectroscopic observations for the GMASS sample were carried out with 
FORS\,2 at the VLT using the blue 300V and the red 300I grisms, at 
resolutions of 440 and 660, respectively. Only a minority of the objects was 
observed in both masks. In view of the particularly long exposure times per 
mask of 11 to 32\,h (and consequently high signal-to-noise ratio - S/N),
the main reason for rejecting a spectrum is the limited wavelength coverage.

For the GOODS-S, a wealth of multiwavelength data is available as part of 
the GOODS legacy (Dickinson et al. \cite{DIC03}; Giavalisco et al. 
\cite{GIA04}). In particular, the deep {\em Spitzer} MIPS observations at 
24\,$\mu$m (Chary et al., in prep.) and 70\,$\mu$m (Frayer et al., in prep.)
are important to probe re-emission by dust at rest-frame mid-IR wavelengths.
This analysis is also based on the available broad-band photometry from
the UV to the near-IR (Kurk et al., in prep.), since it utilises galaxy 
parameters derived from model fits to the photometric data. For instance, 
both the star-formation rate (SFR) and the total stellar mass ($M_{\star}$) 
of a galaxy used here were derived from fitting broad-band SEDs (extending 
from U band to the IRAC 8\,$\mu$m band) using the {\em Hyperz} code of 
Bolzonella et al. (\cite{BOL00}), the redshift being fixed to the measured 
spectroscopic value (cf. Pozzetti et al., in prep.). In particular, fitting 
made use of the stellar population synthesis models of Maraston
(\cite{MARA05})\footnote{We use these particular models because they are the 
only published models including the thermally pulsating asymptotic giant 
branch phase for intermediate-mass ($2 - 5$\,M$_{\odot}$) stars.} including 
a Salpeter (\cite{SAL55}) initial mass function (IMF) and fixed solar 
metallicity ($\mathrm{Z}_{\odot}$), whereas age was allowed to vary up to the 
lookback time of the object. Star-formation histories (SFHs) were described 
by exponentials with $e$-folding times between $100$\,Myr and infinity 
(constant SFR). Dust attenuation was assumed to be described as in Calzetti 
et al. (\cite{CAL00}).

Finally, we note that morphologies of the GMASS galaxies were determined by a 
{\em Galfit} (Peng et al. \cite{PEN02}) analysis of the HST-ACS imaging at 
8500\,\AA{} from the GOODS-S (Cassata et al., in prep.). For $1.5 < z < 2.5$,
the $z$-band maps rest-frame mid-UV to near-UV wavelengths, so that the
morphological parameters reflect the distribution of young stellar
populations. In this study, we adopt measurements of the effective radius
$r_\mathrm{e}$, the S\'ersic (\cite{SER68}) index $n_\mathrm{ser}$, and the
axial ratio $b/a$. Moreover, we investigate the CAS parameters concentration
$C$ and asymmetry $A$ measured according to a method described by Cassata et 
al. (\cite{CAS05}).

\subsection{The FDF, K20, and GDDS samples}\label{othersamples}

The GMASS sample is complemented by the spectroscopic sample investigated by 
\cite{NOL07}, consisting of 108 actively star-forming galaxies at 
$1 < z < 2.5$. Of these 108 objects, 66 galaxies were selected from the 
(mostly) $I$-limited spectroscopic catalogue of the FORS deep field (FDF)
spectroscopic survey (Noll et al. \cite{NOL04}) and originally inspected for 
$z > 2$ by \cite{NOL05}. In addition, 34 galaxies were taken from the 
$K_\mathrm{s} < 20$-selected K20 survey in the GOODS-S and a field around 
quasar 0055-2659 (Cimatti et al. \cite{CIM02}; Mignoli et al. \cite{MIG05}),
and 8 galaxies were selected from the $I$- and $K_\mathrm{s}$-limited
Gemini deep deep survey (GDDS, Abraham et al. \cite{ABR04}). For 88 objects
from the FDF and K20 samples, the UV continuum slope could be determined
in addition to the presence/absence of the UV bump (see \cite{NOL07}).

\subsection{Basic sample properties}\label{basics} 

\begin{figure}
\centering 
\includegraphics[width=8.8cm,clip=true]{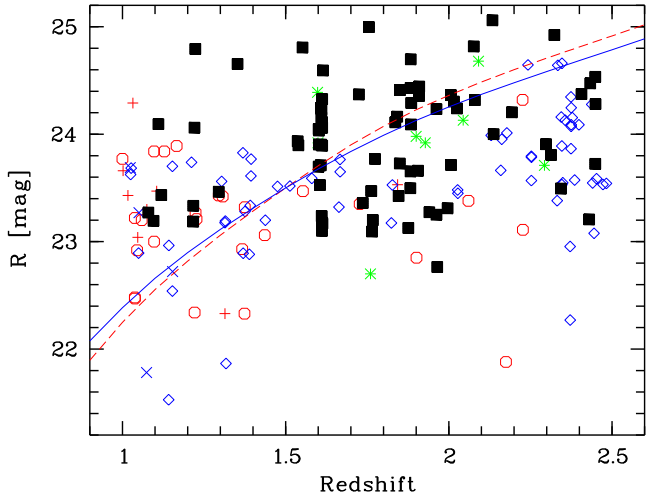}
\caption[]{R-band magnitude versus redshift for the total sample of 186 
actively star-forming galaxies at $1 < z < 2.5$ selected from the GMASS
(filled squares), FDF spectroscopic survey (lozenges and crosses), the K20 
survey (circles and plus signs), and the GDDS (asterisks). Crosses, plus 
signs, and asterisks mark galaxies without a determination of the UV 
continuum slope. To illustrate selection biases, the curves show lines of 
constant luminosity for typical $1 < z < 1.5$ FDF (solid line) and K20 
(dashed line) spectra. Both curves intersect at $R = 23.5$ and $z = 1.5$.}   
\label{fig_R_z}
\end{figure}

\begin{figure}
\centering 
\includegraphics[width=8.8cm,clip=true]{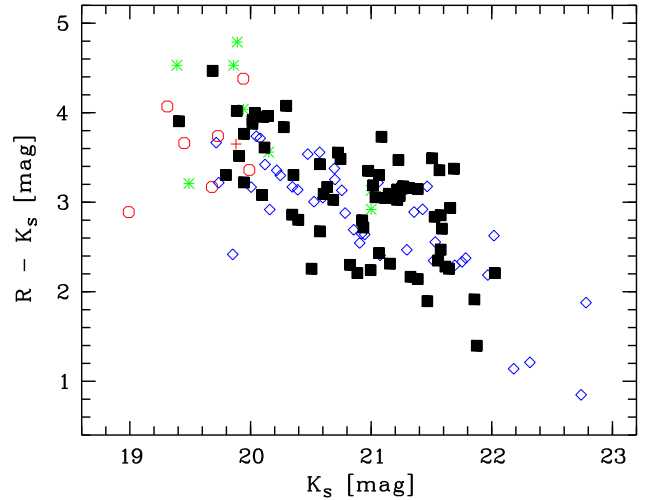}
\caption[]{$R - K_\mathrm{s}$ colour versus $K_\mathrm{s}$ magnitude
for sample galaxies with $1.5 < z < 2.5$. Symbols are the same as in 
Fig.~\ref{fig_R_z}.}   
\label{fig_RK_K}
\end{figure}

Figure~\ref{fig_R_z} shows the range of R-band magnitudes and redshifts 
covered by the different samples considered in this study, including the 10 
GMASS UV-luminous galaxies at $1 < z < 1.5$. All but two \cite{NOL07} objects 
are brighter than $R = 24$ for $z < 2$, and a few reach $\mathrm{R}=24.7$ at
higher redshifts. With GMASS, fainter galaxies ($24 < R < 25$) represent half
of the sample ($\langle R \rangle = 23.9$) and are probed for the entire
redshift range studied. Additional differences in the redshift distributions
exist, as previously anticipated: 76\% of the K20 galaxies are
at $1< z < 1.5$, whereas the FDF sample mainly consists of objects at 
$1 < z < 1.5$ (38\%) or $2 < z < 2.5$ (52\%); in contrast, 67\% of the GMASS 
objects are located at $1.6 < z < 2.1$. As indicated by curves of fixed
luminosity in Fig.~\ref{fig_R_z}, the brightness requirements for 
spectroscopy (and possible evolution in the galaxy population) cause an 
unavoidable increase in the average rest-frame UV luminosity of the sample 
with redshift. Therefore, redshift-dependent results have to be interpreted 
with care.
  
As additional information, Fig.~\ref{fig_RK_K} illustrates the 
$R - K_\mathrm{s}$--$K_\mathrm{s}$ colour--magnitude diagram of the sample 
galaxies at $1.5 < z < 2.5$, where most of the selected GMASS galaxies lie. 
There is good overlap between GMASS and \cite{NOL07} objects at these 
redshifts, the GMASS galaxies being only $0.2$\,mag brighter and $0.3$\,mag 
redder than the coeval FDF galaxies ($K_\mathrm{s} = 20.8$ versus $21.0$, and 
$R - K_\mathrm{s} = 3.1$ versus $2.8$, respectively). Hence, in spite of the 
different selection criteria (i.e., based on the rest-frame near-IR and UV 
domains for the GMASS and FDF samples, respectively), the photometric 
properties of the selected GMASS galaxies are similar to those of the 
selected \cite{NOL07} galaxies. As Fig.~\ref{fig_RK_K} demonstrates, no 
extremely red object (e.g., Pozzetti \& Mannucci \cite{POZ00}) is present in 
the full sample (cf. Fig.~\ref{fig_RK_K}). Aspects of dust attenuation in 
these objects were discussed by Pierini et al. (\cite{PIE04A}) and Pierini et 
al. (\cite{PIE05}). From the results of these analyses, we conclude that our 
sample galaxies are probably neither heavily reddened, young starbursts nor 
normal star-forming, bulge-dominated systems. In principle, the lack of faint 
red galaxies in our sample due to brightness limits could affect the results 
of our analysis. However, investigations of the relation between mass and 
extinction (Greggio et al. \cite{GREG08}) suggest that relatively massive 
star-forming galaxies with such properties could be rare by nature. Hence, we 
can reasonably assume that not detecting such galaxies (if any) does not bias 
our conclusions.

\section{Data analysis}\label{analysis}

In this section, we briefly describe our method for constraining the shape
of the UV extinction curve from optical spectroscopy of high-redshift 
galaxies. In a first step, constraints are obtained from a suitable 
parametric description of the rest-frame UV SED of a galaxy 
(Sect.~\ref{parameters}), as extensively discussed in \cite{NOL05} and 
\cite{NOL07}. In a second, new step, the detailed shape of the observed UV 
bump is derived by fitting a model (Sect.~\ref{lorentz}).

\subsection{Parametric description of the UV continuum}\label{parameters}

\begin{figure}
\centering 
\includegraphics[width=8.8cm,clip=true]{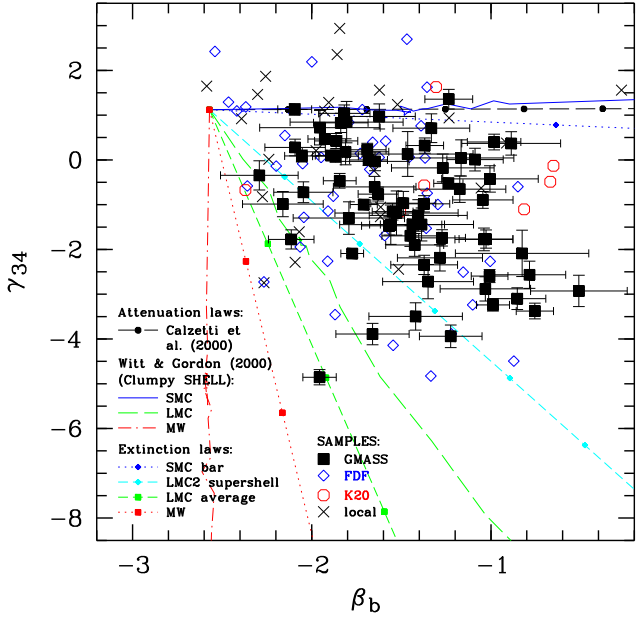}
\caption[]{The UV-bump indicator $\gamma_{34}$ versus the UV reddening measure 
$\beta_{\rm b}$ for the GMASS (filled squares), FDF (open lozenges), and K20 
galaxies (open circles) at $1.5 < z < 2.5$, and a comparison sample of 24 
local starburst galaxies (crosses; see \cite{NOL05}). The diagram also shows 
different dust attenuation models (see legend) for a Maraston (\cite{MARA05}) 
model with Salpeter IMF, continuous SF, an age of 100\,Myr, and solar 
metallicity. The symbols are plotted in intervals of $\Delta E_{B-V} = 0.1$. 
A Maraston model with a SF $e$-folding time of 3\,Gyr and an age of 3\,Gyr 
would produce intrinsically redder continua and, thus, would slightly shift 
the plotted model curves by $0.20$ to the right and by $0.08$ to the top.}
\label{fig_gamma34_betab}
\end{figure}

\cite{NOL05} introduced a parametrisation of the rest-frame UV SED of a 
star-forming, dusty galaxy, which highlights the presence of the UV bump, 
and offers constraints on the far-UV slope of the extinction curve. It is 
based on power-law fits to different sub-regions of the UV continuum, in a
similar way to the parametrisation used by Calzetti et al. (\cite{CAL94}) for 
measuring the UV continuum slope $\beta$. Hereafter, we utilise only the 
parameters $\gamma_{34}$ and $\beta_\mathrm{b}$ (introduced by \cite{NOL07}). 
The first parameter represents the difference between the power-law slopes
$\gamma_3$ and $\gamma_4$ measured in the wavelength ranges 
$1900$--$2175$\,\AA{} and $2175$--$2500$\,\AA{}, respectively, excluding 
strong narrow features. By construction, $\gamma_{34}$ is particularly 
suitable for identifying the presence of the broad absorption feature centred 
on about 2175\,\AA{}: the more negative $\gamma_{34}$, the higher is the 
probability that a UV bump of some strength is present (see 
Fig.~\ref{fig_gamma34_betab}). Moreover, $\gamma_{34}$ depends only very 
little on stellar population properties (\cite{NOL05}). Negative values of
$\gamma_{34}$ are almost impossible for combinations of stellar populations
attenuated by bump-less reddening laws only. \cite{NOL05} chose 
$\gamma_{34} = -2$ as a fiducial threshold for a robust detection of the UV 
bump (see \cite{NOL07} for an updated discussion of this threshold). The 
second parameter $\beta_\mathrm{b}$, measured at $1750$--$2600$\,\AA{}, 
indicates the continuum slope across the UV bump, and is therefore a 
reasonable proxy for the dust reddening of the UV continuum, as well as
$\beta$ (cf. \cite{NOL07}). As in \cite{NOL07}, $\beta_\mathrm{b}$ is
preferred to $\beta$ (measured down to Ly$\alpha$) because the latter cannot
be derived from optical spectra for galaxies at $z < 2$. By using
$\beta_\mathrm{b}$ for all galaxies across the entire redshift range
$1.0$--$2.5$, we are able to compare their properties in a consistent way.

We know that $\gamma_{34}$ alone does not infer whether the extinction curve 
is similar to either the MW or LMC curve, because it indicates the 
{\em apparent} amplitude/area of the UV bump, which is a function of the dust 
opacity (see, e.g., Witt \& Gordon \cite{WIT00}). Since the dust reddening of 
UV SEDs depends on the opacity as well, it is the distribution of a galaxy in 
the $\beta_\mathrm{b}$--$\gamma_{34}$ plane that identifies the extinction 
curve (see Fig.~\ref{fig_gamma34_betab}).

\subsection{Lorentzian-like fitting of the UV bump}\label{lorentz}

The parameter $\gamma_{34}$ can easily be measured in spectra with a 
relatively narrow spectral coverage, and even in noisy ones, but it loosely 
constrains the observed shape of the UV bump, and, thus, the kind of 
extinction curve (i.e., MW-, LMC-, or SMC-like). The high S/N and wavelength 
coverage of the GMASS spectra offer the opportunity for a more accurate 
description of the UV bump. In particular, for $z \sim 2$ the UV bump sits 
almost in the middle of the observed wavelength range, which allows the bump 
to be fitted in its shallowest wings. For optimum results, we investigate 
stacked spectra (introducing subsampling) and not individual ones (see 
Sect.~\ref{shape}).

According to Fitzpatrick \& Massa (\cite{FIT90}, FM), the shape of the UV 
bump present in extinction curves is most accurately described by a 
Lorentzian-like ``Drude'' profile 
\begin{equation}\label{eq_drude}
D(x,x_0,\gamma) = \frac{x^2}{(x^2 - {x_0}^2)^2 + x^2\gamma^2},
\end{equation} 
where $x$, $x_0$, and $\gamma$ are expressed in units of inverse wavelength,
$x_0$ being the central wavelength and $\gamma$ the inverse of the FWHM
of the UV bump. In the FM parametrisation, extinction curves $k(\lambda - V)$ 
are normalised by the reddening $E_{B-V}$, and the UV bump is reproduced by 
$c_3\,D(x,x_0,\gamma)$, where $c_3$ is an amplitude (in units of inverse 
wavelength in quadrature). In later work, Fitzpatrick \& Massa (\cite{FIT07}) 
defined two additional parameters, based on combinations of the basic 
parameters $x_0$, $\gamma$, and $c_3$. In particular, 
$E_\mathrm{bump} \equiv c_3 / \gamma^2$ is the maximum height above the 
linear baseline extinction, and $A_\mathrm{bump} \equiv \pi c_3 / (2\gamma)$
indicates the area subtended by the UV bump.

In analogy with the FM formalism, we characterise the observed UV-bump in 
galaxy spectra as an excess with respect to a baseline attenuation. At 
variance with extinction curves measured for sightlines to individual stars,
the observed UV bump in a galaxy spectrum is affected by radiative transfer
effects since spectra probe large regions of galaxies (see, e.g., Ferrara et 
al. \cite{FER99}; Pierini et al. \cite{PIE04B} in case of photometry). We set
the baseline to be the attenuation curve of nearby starburst galaxies
(Calzetti et al. \cite{CAL00}), where the wavelength dependence is fixed and
the amount of attenuation at a given wavelength scales with the {\em observed}
reddening $E_{B-V}$ at optical wavelengths. Since the curvature of the
selected reference curve for the continuum attenuation is not supposed to
differ strongly from the unknown real shape within the range of the UV bump,
the choice of the baseline mostly affects the measured amplitudes of the UV
bump, while centre and width are expected to be almost invariable. Our
approach is unorthodox but allows us to link our results to those obtained
from studies of galaxy evolution, which adopt the Calzetti law as a
description of dust attenuation in galaxies at cosmological distances.
Therefore, we refer to the parameters derived for our sample galaxies as
``FM-like''.

In our procedure, regions of the spectra that exclude the 2175\,\AA{} 
feature\footnote{A region across the UV bump of width $900$--$1000$\,\AA{} is 
masked to avoid even faint wings.} and other known strong spectral lines
are fitted by different Maraston (\cite{MARA05}) models with exponentially 
decaying star-formation rates and reddened by a Calzetti et al. 
(\cite{CAL00}) attenuation law. By doing so, we are able to interpolate 
across the UV bump. Taking into account uncertainties and approximations, 
fits are considered to be ``good'' if their significance reaches at least 
95\% of that of the best-fit model. For these good fits, the residual 
2175\,\AA{} feature (if any) is detectable in a spectrum created by obtaining
the ratio of the initial spectrum to the model. Retrieving the properties of 
the observed UV bump therefore depends only on the best-fit $E_{B-V}$ value 
of the Calzetti law. The uncertainty in $E_{B-V}$ mainly affects the observed 
UV-bump amplitude $c_3$, whereas $x_0$ and $\gamma$ are less dependent on the 
best-fit model. These FM-like parameters are derived from Lorentzian fits to 
each continuum-subtracted UV bump. Lorentzian fits are considered as ``good'' 
if their significance is at least 95\% of that of the best-fit model. They 
are used to compute the mean and standard deviation of each FM-like 
parameter. To estimate the influence of noise on these fits, a set of 
synthetic spectra modified on the basis of the error function are computed 
and analysed in the same way as the real data. The final variance in the 
FM-like parameters measured for a stacked spectrum is given by the standard 
deviation of the good fits and the statistical noise, added in quadrature.

\section{Results}\label{results}

We first present and discuss results obtained from the parametrisation and 
modelling of the rest-frame UV continua of our high-redshift galaxies
(Sect.~\ref{frequency}), whereas detailed shapes of the observed UV bumps
are illustrated in Sect.~\ref{shape}. The link between carriers of the UV 
bump and properties of the parent galaxies is addressed in 
Sect.~\ref{galaxyprop}. Finally, the existence of a relation between the 
presence of these carriers and the amount of re-emission by dust at 
rest-frame 8\,$\mu$m is discussed in Sect.~\ref{UV_IR}.

\subsection{Extinction curves with a UV bump are not unusual among 
high-redshift galaxies with gas and dust}
\label{frequency}

\begin{figure}
\centering 
\includegraphics[width=8.8cm,clip=true]{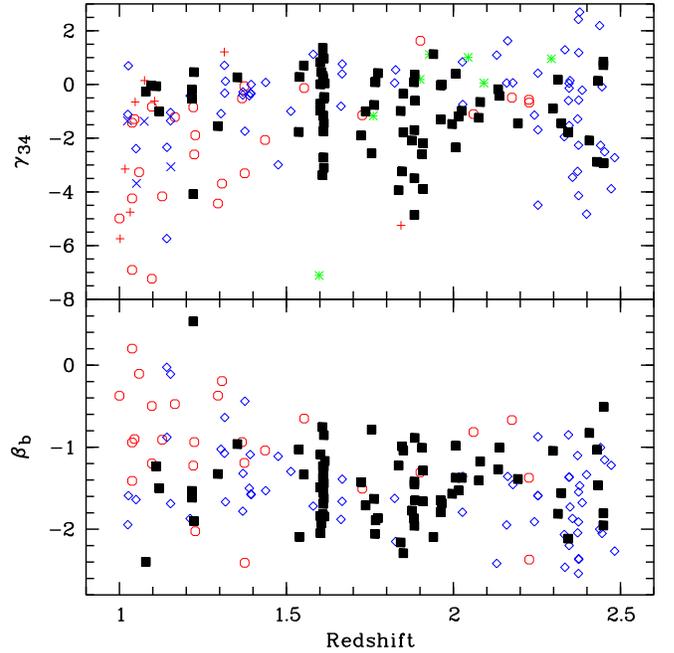}
\caption[]{The UV-bump indicator $\gamma_{34}$ ({\em upper panel}) and the UV 
reddening measure $\beta_{\rm b}$ ({\em lower panel}) versus redshift for 78 
GMASS galaxies at $1 < z < 2.5$ (filled squares) in comparison to 108 
galaxies from \cite{NOL07} (20 without $\beta_\mathrm{b}$ measurement). 
Symbols are the same as in Fig.~\ref{fig_R_z}.}  
\label{fig_gamma34_betab_z}
\end{figure}

\begin{figure}
\centering 
\includegraphics[width=8.8cm,clip=true]{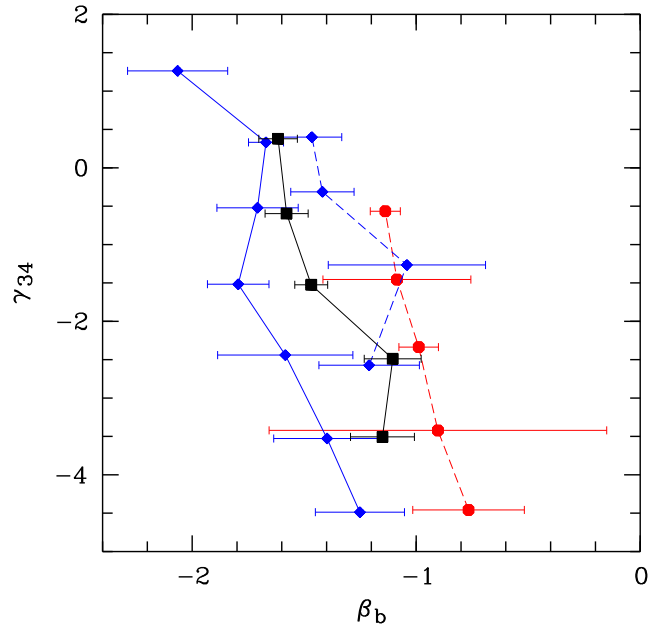}
\caption[]{Average UV reddening measure $\beta_{\rm b}$ for unit bins of the 
UV-bump indicator $\gamma_{34}$. The following subsamples are shown: GMASS 
for $1.5 < z < 2.5$ (squares and solid lines), FDF for $1.5 < z < 2.5$
(lozenges and solid lines), FDF for $1 < z < 1.5$ (lozenges and dashed 
lines), and K20 for $1 < z < 1.5$ (circles and dashed lines). Only bins
with at least two data points are plotted. The $\beta_{\rm b}$ uncertainties 
shown are mean errors derived from the sample variance and the uncertainties
of individual measurements.}
\label{fig_gamma34_bins}
\end{figure}

Figure~\ref{fig_gamma34_betab_z} illustrates the redshift distribution of the 
proxy for the presence of a UV bump in spectra, $\gamma_{34}$, and the 
continuum slope in the UV-bump region, $\beta_\mathrm{b}$, for the GMASS, 
FDF, K20, and GDDS samples. A total of 186 actively star-forming, 
intermediate-mass galaxies at $1 < z < 2.5$ is shown, including the 10 GMASS 
objects at $1 < z < 1.5$. As from \cite{NOL07}, the fraction of UV-luminous 
galaxies at these redshifts that clearly exhibit a 2175\,\AA{} feature in 
their spectra (i.e., $\gamma_{34} < -2$) is about 30\%. In particular, this 
fraction is 25\% for the combined FDF and K20 samples at $1 < z < 1.5$, and 
36\% for the FDF sample at $2.2 < z < 2.5$. A similar fraction of 24\% 
{\em s}ignificant {\em U}V {\em b}ump {\em g}alaxies (SUBGs) characterises the 
GMASS sample at $1.5 < z < 2.5$, 13 out of 17 detections being for objects 
at $1.5 < z < 2.2$. This fraction drops to 10\% for the GMASS galaxies at 
$1 < z < 1.5$.

We verified that, for the GMASS sample (see Sect.~\ref{sample}), subsampling 
in terms of rest-frame UV luminosity (i.e., splitting the sample at 
$R = 24.0$) or large scale structure (see the overdensity of galaxies at 
$z \sim 1.61$; Kurk et al. \cite{KUR08B}) does not impact this fraction. 
Hence, we conclude that the non-detection of SUBGs at $1.5 < z < 2.5$ in the 
FDF and K20 samples (see \cite{NOL07}) is probably due to low number 
statistics, plus possible selection and detection biases. In support of this 
conclusion, we note that the distributions of the GMASS, FDF, and K20 
galaxies at $1.5 < z < 2.5$ in the $\beta_\mathrm{b}$--$\gamma_{34}$ 
diagnostic diagram are similar (Fig.~\ref{fig_gamma34_betab}). In particular, 
the average value of $\gamma_{34}$ for the GMASS galaxies at $1.5 < z < 2.5$ 
is consistent with that of coeval FDF galaxies ($-1.01 \pm 0.17$ versus 
$-0.70 \pm 0.30$) within the errors. The average value of the reddening in 
the UV, $\beta_\mathrm{b}$, is slightly higher ($-1.48 \pm 0.05$ versus 
$-1.71 \pm 0.07$). At $1 < z < 1.5$, the combined FDF and K20 samples 
exhibit stronger 2175\,\AA{} features 
($\langle \gamma_{34} \rangle = -1.23 \pm 0.23$) and redder UV continua
($\langle \beta_\mathrm{b} \rangle = -1.23 \pm 0.08$) due to a population of 
galaxies with extreme UV-continuum properties, which is not present in our 
sample at higher redshifts (see Fig.~\ref{fig_gamma34_betab_z}). 
 
The GMASS sample confirms that more negative values of $\gamma_{34}$ are 
found as $\beta_\mathrm{b}$ increases. The galaxy distribution in 
Fig.~\ref{fig_gamma34_betab} is interpreted by making use of different 
dust-reddened stellar-population models. In particular, for a 100\,Myr-old,
solar metallicity, continuous star-formation model of Maraston
(\cite{MARA05}), we show models with increasing $E_{B-V}$ for
the attenuation law of Calzetti et al. (\cite{CAL00}), different clumpy
shell models of Witt \& Gordon (\cite{WIT00}), and uniform screen models
using different extinction curves obtained for the SMC, LMC (Gordon et al.
\cite{GOR03}), and the MW (Cardelli et al. \cite{CAR89}). The distribution
of the data points is well constrained by these models, and evidence of 
extinction curves with a form other than that of the SMC (with no UV bump) is 
clear. The evidence of non-SMC extinction curves is insensitive to the choice 
of the stellar-population model. For instance, considering a Maraston model 
with a SFR $e$-folding time of 3\,Gyr at an epoch of 3\,Gyr after the 
beginning of the SF activity would shift the plotted model curves only by 
$+0.20$ in $\beta_\mathrm{b}$ and $+0.08$ in $\gamma_{34}$.

According to the models shown in Fig.~\ref{fig_gamma34_betab}, the presence 
of a moderate UV bump (as in the LMC extinction curve) is suggested. 
However, the identification of the extinction curve becomes complex, if 
models implementing age-dependent extinction scenarios are considered.
In fact, as a given simple stellar population becomes less attenuated as time 
goes by (i.e., as the average age of the stars increases), extinction 
curves with more pronounced UV bumps (e.g., as for the MW) are allowed by the 
distribution of the data points (see \cite{NOL07} for a detailed 
discussion). This scenario can be implemented by, e.g., realistic dust/star 
distributions similar to those of nearby disc galaxies (e.g., Silva et al. 
\cite{SIL98}; Pierini et al. \cite{PIE04B}). However, it cannot explain the 
existence of spectra with extremely negative values of $\gamma_{34}$. In this
case, extraplanar dust must be invoked (\cite{NOL07}). Hence, time-independent 
screen models (without scattering) or dusty shells appear to describe dust 
attenuation well in the high-redshift galaxies of our sample, and, thus, the 
data distribution in Fig.~\ref{fig_gamma34_betab}.

Besides the overall agreement, some slight but interesting differences 
between samples exist, as suggested by Fig.~\ref{fig_gamma34_bins}, which 
reproduces average values of $\beta_\mathrm{b}$ for $\Delta\gamma_{34} = 1$ 
bins. In fact, moving from the FDF galaxies at $1.5 < z < 2.5$ 
($\langle z \rangle = 2.21$) to the GMASS galaxies in the same redshift range  
($\langle z \rangle = 1.89$), and, then, from the FDF galaxies at 
$1 < z < 1.5$ ($\langle z \rangle = 1.27$) to the K20 galaxies in the same 
redshift range ($\langle z \rangle = 1.18$), $\beta_\mathrm{b}$ tends to 
increase or the rest-frame UV continua become redder. Comparison of values
obtained for the different samples shows that the significance of this trend 
ranges between 1 and $3\,\sigma$. The most reliable values are found for the 
largest redshift differences. In principle, differences in the selection 
criteria, average luminosities, and redshifts (partly driven by the small 
number statistics available for individual samples) or systematic
errors\footnote{The FDF galaxies dominate the sample of galaxies beyond
$z = 2.3$. Their measured values of $\beta_\mathrm{b}$ can be affected by
systematic errors due to the presence of strong OH residuals in the red
optical wavelength region, where the redshifted $\beta_\mathrm{b}$ was
measured.} could play a role. For the GMASS galaxies at $1.5 < z < 2.5$, 
we checked that the variation in $\beta_\mathrm{b}$ is negligible 
($0.06 \pm 0.09$) for objects brighter or fainter than $R = 24$. Therefore, 
the presence of redder UV continua as redshift decreases may have a 
physical origin, i.e., ageing of the overall stellar population and/or an 
increase in the overall ISM opacity. A change in the dust composition and/or
distribution is possible as well.

Interestingly, the sample of 24 local starburst galaxies observed by IUE
and selected by \cite{NOL05} exhibits relatively bluer UV 
continua than our sample galaxies:
$\langle\beta_\mathrm{b}\rangle = -1.79 \pm 0.11$
for $\langle\gamma_{34}\rangle = 0.27 \pm 0.32$.
Suitable models of a MW-like galaxy that involve stellar population
synthesis and radiative transfer for a disc geometry (see Pierini et al. 
\cite{PIE03}) predict redder UV continua than starbursts in the local 
Universe. According to the models used by \cite{NOL05}, ageing of the 
stellar populations would however produce only a moderate shift in 
$\beta_\mathrm{b}$ equal to $0.1$--$0.2$. Unfortunately, a robust age-dating 
of the stellar populations of all our sample galaxies is not possible, since 
the redshifted Balmer and 4000\,\AA{} breaks are detectable in optical
spectra only for $z < 1.5$ (see \cite{NOL07}), and model fitting 
provides quite uncertain age estimates.

\subsection{Constraining the attenuation curve}\label{shape}

\begin{figure}
\centering 
\includegraphics[width=8.8cm,clip=true]{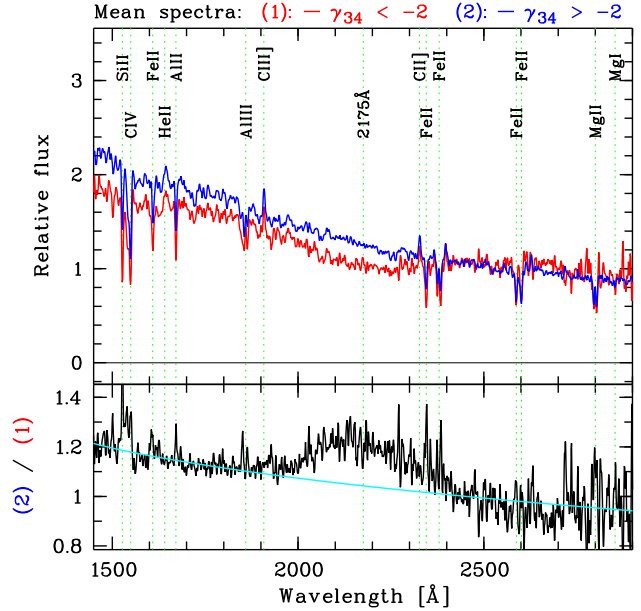}
\caption[]{Comparison of composite spectra of $1.5 < z < 2.5$ GMASS galaxies 
with $\gamma_{34} > -2$ (blue) and $\gamma_{34} < -2$ (red), respectively. 
The lower panel gives the ratio of both composites, normalised at 
$2400 - 2570$\,\AA{}. Deviations from a Calzetti law in the UV-bump range are 
illustrated by the overplotted best-fit Calzetti attenuation curve with 
$E_{B-V} = 0.08$.}
\label{fig_compmean}
\end{figure}

\begin{table}
\caption[]{Average properties of the three GMASS subsamples at 
$1.5 < z < 2.5$}
\label{tab_avpar}
\centering
\begin{tabular}{l c @{ \ \ \ } c @{ \ \ \ } c}
\hline\hline
\noalign{\smallskip}
Par. & 
$\gamma_{34} > -2$        & $\gamma_{34} > -2$        & $\gamma_{34} < -2$ \\
     &
$\beta_\mathrm{b} < -1.5$ & $\beta_\mathrm{b} > -1.5$ &                    \\ 
\noalign{\smallskip}
\hline
\noalign{\smallskip}
$N$                                       & 
29                & 23                & 16                \\
$z$                                       &
$1.86 \pm 0.05$   & $1.85 \pm 0.05$   & $1.91 \pm 0.06$   \\ 
$R$                                       &
$23.74 \pm 0.09$  & $24.20 \pm 0.10$  & $23.99 \pm 0.13$  \\
$K_\mathrm{s}$                            &
$21.09 \pm 0.11$  & $20.74 \pm 0.15$  & $20.60 \pm 0.13$  \\ 
$\gamma_{34}$                             &
$-0.20 \pm 0.16$  & $-0.62 \pm 0.19$  & $-3.02 \pm 0.19$  \\
$\beta_\mathrm{b}$                        &
$-1.83 \pm 0.04$  & $-1.24 \pm 0.04$  & $-1.18 \pm 0.10$  \\ 
$\log L_{1500}$ [W/\AA{}]                 &
$33.95 \pm 0.04$  & $33.68 \pm 0.03$  & $33.82 \pm 0.07$  \\ 
$\log L_{2500}$ [W/\AA{}]                 &
$33.61 \pm 0.04$  & $33.46 \pm 0.04$  & $33.60 \pm 0.06$  \\
$E_{B-V}$                                 &
$0.133 \pm 0.020$ & $0.244 \pm 0.022$ & $0.304 \pm 0.020$ \\
$\log L_\mathrm{bol}$ [L$_{\odot}$]       &
$11.37 \pm 0.08$  & $11.56 \pm 0.07$  & $11.85 \pm 0.04$  \\
$\log \mathrm{SFR}$ [M$_{\odot}$/yr]      &
$1.53 \pm 0.09$   & $1.70 \pm 0.08$   & $2.05 \pm 0.04$   \\
$\log M_{\star}$ [M$_{\odot}$]            &
$9.94 \pm 0.05$   & $10.12 \pm 0.08$  & $10.15 \pm 0.07$  \\
$\log \phi$ [Gyr$^{-1}$]                  &
$0.52 \pm 0.08$   & $0.51 \pm 0.11$   & $0.85 \pm 0.07$   \\
$r_\mathrm{e}$ [kpc]$^\mathrm{a}$         &
$2.84 \pm 0.48$   & $2.65 \pm 0.37$   & $2.52 \pm 0.38$   \\ 
$n_\mathrm{ser}$                          &
$1.56 \pm 0.32$   & $1.79 \pm 0.44$   & $0.77 \pm 0.15$   \\
$b/a$                                     &
$0.46 \pm 0.03$   & $0.54 \pm 0.04$   & $0.46 \pm 0.04$   \\
$C_\mathrm{CAS}$                          &
$2.57 \pm 0.07$   & $2.63 \pm 0.07$   & $2.55 \pm 0.08$   \\ 
$A_\mathrm{CAS}$                          &
$0.40 \pm 0.03$   & $0.42 \pm 0.03$   & $0.43 \pm 0.03$   \\
$f_{24\,\mu\mathrm{m}}$ [$\mu$Jy]         &
$37.3 \pm 14.8$   & $39.1 \pm 10.4$   & $63.3 \pm 9.9$    \\ 
$\log L_{8\,\mu\mathrm{m}}$ [L$_{\odot}$] &
$10.28 \pm 0.07$  & $10.35 \pm 0.09$  & $10.73 \pm 0.07$  \\
$\log L_\mathrm{IR}$ [L$_{\odot}$]        &
$11.11 \pm 0.09$  & $11.21 \pm 0.11$  & $11.66 \pm 0.08$  \\
$\log (L_\mathrm{IR} / L_\mathrm{UV})$    &
$0.50 \pm 0.10$   & $0.89 \pm 0.13$   & $1.25 \pm 0.12$   \\
\noalign{\smallskip}
\hline
\end{tabular}
\begin{list}{}{}
\item[$^\mathrm{a}$] from $\log r_{\rm e}$
\end{list}
\end{table}

\begin{table*}
\caption[]{UV-bump parameters analogous to those introduced by Fitzpatrick \& 
Massa (\cite{FIT90})}
\label{tab_fmpar}
\centering
\begin{tabular}{@{ \ } l l c @{ \ \ \ } c @{ \ \ \ } c @{ \ \ } c @{ \ } c 
@{ \ }}
\hline\hline
\noalign{\smallskip}
Sample$^\mathrm{a}$ & Selection & $E_{B-V}$ & $x_0$ [$\mu$m$^{-1}$] & 
$\lambda_0$ [\AA{}] & $\gamma$ [$\mu$m$^{-1}$] & FWHM [\AA{}] \\
\noalign{\smallskip}
\hline
\noalign{\smallskip}
(a) gm\_wbb &
GMASS, $1.5 < z < 2.5$, $\gamma_{34} > -2$, $\beta_{\rm b} < -1.5$ &
$0.164 \pm 0.069$ & $4.493 \pm 0.024$ & $2226 \pm 12$ &  $0.498 \pm 0.113$ & 
$247 \pm 56$ \\ 
(b) gm\_wbr &
GMASS, $1.5 < z < 2.5$, $\gamma_{34} > -2$, $\beta_{\rm b} > -1.5$ &
$0.286 \pm 0.071$ & $4.558 \pm 0.024$ & $2194 \pm 11$ & $0.682 \pm 0.052$ & 
$328 \pm 25$ \\
(c) gm\_sb &
GMASS, $1.5 < z < 2.5$, $\gamma_{34} < -2$ & $0.296 \pm 0.069$ &
$4.572 \pm 0.010$ & $2187 \pm 05$ & $0.573 \pm 0.021$ & $274 \pm 10$ \\
(d) n07\_sb &
FDF/K20, $1.0 < z < 1.5$, $\gamma_{34} < -2$ & $0.344 \pm 0.078$ & 
$4.570 \pm 0.019$ & $2188 \pm 09$ & $0.583 \pm 0.083$ & $279 \pm 38$ \\
(e) LMC2\_low$\gamma$ & LMC\,2 supershell ($\gamma < 1$) & 
$0.188 \pm 0.009$ & $4.540 \pm 0.053$ & $2203 \pm 26$ & $0.684 \pm 0.042$ & 
$332 \pm 21$ \\
\noalign{\smallskip}
\hline
\noalign{\smallskip}
Sample$^\mathrm{a}$ & Selection & $c_3$ [$\mu$m$^{-2}$] & $E_{\rm bump}$ & 
$A_{\rm bump}$ [$\mu$m$^{-1}$] & & \\
\noalign{\smallskip}
\hline
\noalign{\smallskip}
(a) gm\_wbb & 
GMASS, $1.5 < z < 2.5$, $\gamma_{34} > -2$, $\beta_{\rm b} < -1.5$ &
$0.116 \pm 0.055$ & $0.467 \pm 0.267$ & $0.366 \pm 0.192$ & & \\ 
(b) gm\_wbr & 
GMASS, $1.5 < z < 2.5$, $\gamma_{34} > -2$, $\beta_{\rm b} > -1.5$ &
$0.262 \pm 0.040$ & $0.562 \pm 0.106$ & $0.602 \pm 0.104$ & & \\
(c) gm\_sb & 
GMASS, $1.5 < z < 2.5$, $\gamma_{34} < -2$ & $0.305 \pm 0.020$ & 
$0.928 \pm 0.077$ & $0.835 \pm 0.062$ & & \\
(d) n07\_sb & 
FDF/K20, $1.0 < z < 1.5$, $\gamma_{34} < -2$ & $0.359 \pm 0.121$ & 
$1.056 \pm 0.186$ & $0.968 \pm 0.251$ & & \\
(e) LMC2\_low$\gamma$ & LMC\,2 supershell ($\gamma < 1$) & 
$0.890 \pm 0.128$ & $1.903 \pm 0.320$ & $2.044 \pm 0.320$ & & \\
\noalign{\smallskip}
\hline
\end{tabular}
\begin{list}{}{}
\item[$^\mathrm{a}$] For the samples at high redshift, the given errors are 
based on the uncertainties in the continuum of the mean spectrum without the 
2175\,\AA{} feature, $E_{B-V}$, and the FM-like parameter fit. For the LMC, 
they represent the sample variance.
\end{list}
\end{table*}

\begin{figure}
\centering 
\includegraphics[width=8.8cm,clip=true]{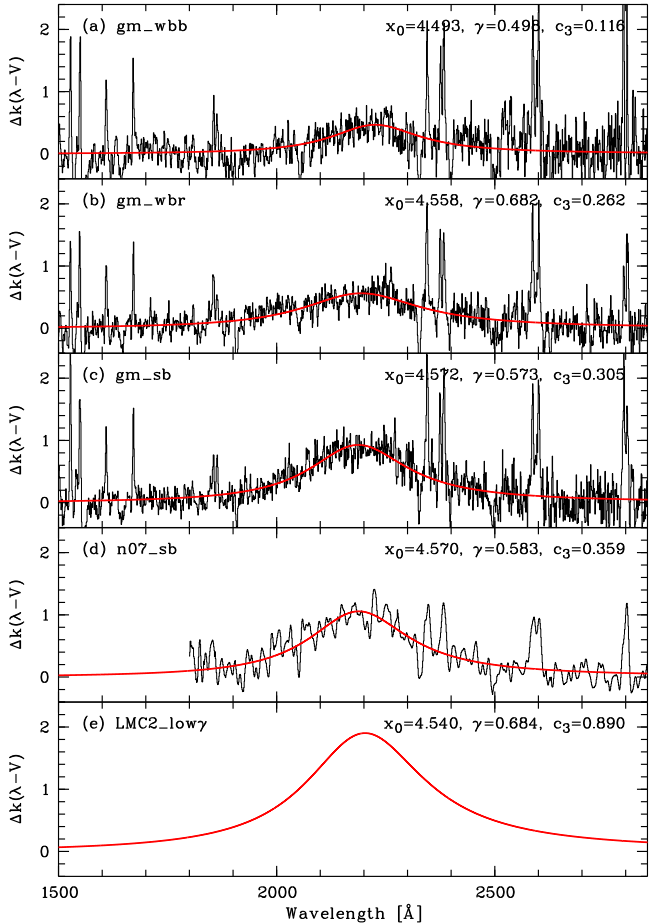}
\caption[]{Best Lorentzian fits of the bump component of the average 
extinction curves of the three GMASS subsamples at $1.5 < z < 2.5$ (a - c) 
and a subsample of FDF and K20 galaxies at $1.0 < z < 1.5$ with 
$\gamma_{34} < -2$ (d). Moreover, the average 2175\,\AA{} bump of a sample of 
measurements in the LMC\,2 supershell region close to 30\,Dor showing low 
$\gamma$ is presented (e). More details about the samples and the fits can be 
found in Table~\ref{tab_fmpar}.}
\label{fig_bumpfits}
\end{figure}

\begin{figure}
\centering 
\includegraphics[width=8.8cm,clip=true]{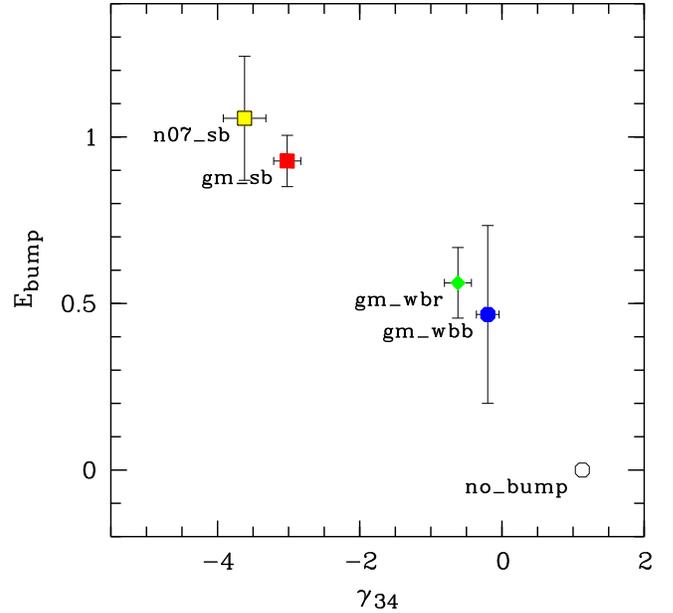}
\caption[]{Amplitude of the UV bump in the extinction curve $E_{\rm bump}$ 
(see Sect.~\ref{lorentz}) versus $\gamma_{34}$ indicating the strength of the 
2175\,\AA{} feature in the spectrum for the three subsamples of GMASS 
galaxies at $1.5 < z < 2.5$ and the \cite{NOL07} subsample of $1 < z < 1.5$ 
galaxies with $\gamma_{34}$. Labels are the same as in Fig.~\ref{fig_bumpfits} 
and Table~\ref{tab_fmpar} but for an open circle, which marks the position of 
an extinction curve without any UV bump. The corresponding $\gamma_{34}$ is 
given for the stellar population parameters described in 
Fig.~\ref{fig_gamma34_betab}.}
\label{fig_Ebump_gamma34}
\end{figure}

\begin{figure}
\centering 
\includegraphics[width=8.8cm,clip=true]{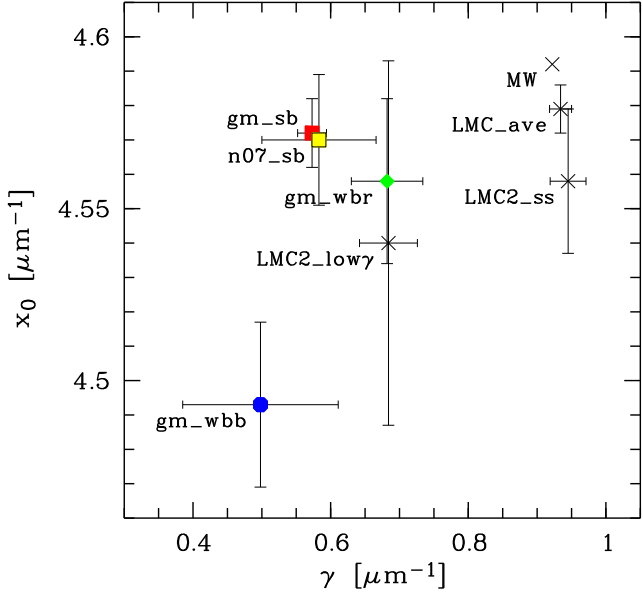}
\caption[]{FM-like parameters $x_0$ (the inverse central wavelenth of the UV 
bump) and $\gamma$ (linked to the inverse FWHM of the UV bump) for the three 
subsamples of GMASS galaxies at $1.5 < z < 2.5$ and the \cite{NOL07} 
subsample of $1 < z < 1.5$ galaxies with $\gamma_{34} < -2$ (filled symbols).
Labels are the same as in Fig.~\ref{fig_bumpfits} and Table~\ref{tab_fmpar}. 
Moreover, crosses mark average FM parameters for the Milky Way (Fitzpatrick 
\& Massa \cite{FIT07}) and different LMC environments (Gordon et al. 
\cite{GOR03}). The latter are the average LMC (`LMC\_ave'), the entire LMC\,2
supershell region (`LMC2\_ss'), and the LMC\,2 supershell region restricted
to stars with $\gamma < 1$\,$\mu$m$^{-1}$ (`LMC2\_low$\gamma$'). The errors
given are measurement uncertainties of the FM parameters for the corresponding 
mean extinction curves except for `LMC2\_low$\gamma$', for which sample mean 
errors are provided.}
\label{fig_x0_gamma}
\end{figure}

In analogy with \cite{NOL05} and \cite{NOL07}\footnote{The \cite{NOL07} 
sample of galaxies at $1 < z < 1.5$ is divided into a blue 
($\beta_\mathrm{b} < -1.5$) and two red subsamples differing in 
$\gamma_{34}$.}, we divide the GMASS sample of galaxies at $1.5 < z < 2.5$ 
into subsamples, using the same thresholds of $\beta_\mathrm{b} = -1.5$ and 
$\gamma_{34} = -2$. In particular, we consider one GMASS subsample with 
$\gamma_{34} < -2$, and two with $\gamma_{34} > -2$ that differ in terms of 
$\beta_\mathrm{b}$. For $1.5 < z < 2.5$, 16 GMASS galaxies are SUBGs, 
whereas 23 and 29 GMASS galaxies have $\gamma_{34} > -2$ plus red or blue UV 
continua, respectively. Average values of $\gamma_{34}$ and 
$\beta_\mathrm{b}$ are listed in Table~\ref{tab_avpar} for these GMASS 
subsamples.

Figure~\ref{fig_compmean} shows a comparison of the composite spectra of the 
GMASS galaxies at $1.5 < z < 2.5$ exhibiting either weak evidence
(at most) of a broad absorption feature centred on 2175\,\AA{} in their 
spectra ($\gamma_{34} > -2$), or a secure one ($\gamma_{34} < -2$). The 
presence of this feature in the second (more reddened) composite spectrum is 
evident by eye.

\subsubsection{The shape of the UV bump}\label{bumpshape}

Lorentzian-like fits are applied to the stacked rest-frame UV continua
obtained from galaxies at $1.5 < z < 2.5$ in the three GMASS subsamples,
and from FDF and K20 galaxies at $1.0 < z < 1.5$ with 
$\gamma_{34} < -2$\footnote{Their composite spectrum is cut at 1800\,\AA{}, 
since only a few spectra extend shortwards. Possible effects of the UV bump 
on the 1800--1900\,\AA{} region have been considered.}, as described in 
Sect.~\ref{frequency}. The resulting FM-like parameters describing the 
morphology of the observed UV bump ---i.e., centre $x_0$ (or $\lambda_0$), 
$\gamma$, and amplitude $c_3$--- and their errors are listed in 
Table~\ref{tab_fmpar}, with the maximum height $E_\mathrm{bump}$ and 
area $A_\mathrm{bump}$ of the observed UV bump, and the mean $E_{B-V}$ of the 
best-fit Calzetti et al. (\cite{CAL00}) attenuation law used as a baseline. 
We note that the values of $E_{B-V}$ obtained from this Lorentzian-like 
fitting are consistent with those obtained from the SED fitting of GMASS 
galaxies (see Sect.~\ref{gmass}) within the errors. However, we acknowledge 
that differences in $E_{B-V}$ become higher for red galaxies with some 
signature of the UV bump in their spectra.

Figure~\ref{fig_bumpfits} shows the quality of the Lorentzian-like best-fit 
functions of the observed (residual) UV bumps obtained for the four 
subsamples. We note that radiative transfer effects only smooth out the UV 
bump in disc configurations (cf. Ferrara et al. \cite{FER99}; Pierini et al. 
\cite{PIE04B}). Furthermore, no significant variation in the width of the 
observed UV bump is found for the different radiative transfer models of  
Witt \& Gordon (\cite{WIT00}) for a given dust composition. Thus, we can 
assume that the residual UV bumps determined from the composite spectra of 
our galaxies are closely related to those present in the extinction curve 
characterising their dust. As a first result, Fig.~\ref{fig_Ebump_gamma34} 
indicates a comfortably good correlation between the maximum height of the UV 
bump $E_\mathrm{bump}$ (dependent on the FWHM and amplitude) and the 
independently measured $\gamma_{34}$. Since the increase in $E_{B-V}$ with 
decreasing $\gamma_{34}$ in our sample (see Table~\ref{tab_fmpar}) is 
obviously unable to remove this correlation, the increasingly negative values 
of $\gamma_{34}$ are indeed associated with more pronounced UV bumps.  

The central wavelengths and widths of the residual UV bumps obtained for
GMASS SUBGs at $1.5 < z < 2.5$ and FDF and K20 SUBGs at $1.0 < z < 1.5$ are 
almost identical. They have counterparts among those measured along different 
sightlines towards the MW and LMC\footnote{The average values of the FM 
parameters for the MW (Fitzpatrick \& Massa \cite{FIT07}) and LMC (Gordon et 
al. \cite{GOR03}) are: 2180\,\AA{} (central wavelength), 440\,\AA{} (FWHM) or 
$0.93$\,$\mu$m$^{-1}$ ($\gamma$), and $3.0$ or $2.7$ ($c_3$), respectively.}
(see Fig.~\ref{fig_x0_gamma}). In particular, the most suitable local 
counterpart is the supershell region LMC\,2 close to 30\,Dor, for which the 
average UV bump from different sightlines (excluding two outliers) is centred 
on $2203 \pm 26$\,\AA{} and has a FWHM equal to $332 \pm 21$\,\AA{}
(cf. Gordon et al. \cite{GOR03}). For instance, the central wavelength
and FWHM of the residual UV bump obtained for the composite spectrum
of GMASS SUBGs are equal to $2187 \pm 5$\,\AA{} and $274 \pm 10$\,\AA{}, 
respectively. In general, the observed UV bumps of our high-redshift galaxies 
are relatively narrow, since they exhibit only about 60\% of the typical 
widths measured for sightlines towards the MW and LMC.

In addition, we note that the two composite spectra obtained for either the 
blue or red GMASS galaxies at $1.5 < z < 2.5$ with no strong evidence
of a UV bump in their individual spectra (i.e., with $\gamma_{34} > -2$)
nevertheless exhibit a weak one. In either case, the central wavelength and 
width of the residual UV bump are not dissimilar to those obtained for 
galaxies with $\gamma_{34} < -2$. This tells us that some traceable dust 
similar to that in the supershell region LMC\,2 seems to be present also
in a non-negligible fraction of the blue and red GMASS non-SUBGs at 
$1.5 < z < 2.5$.

\subsubsection{The shape of the attenuation curve}\label{curveshape}

Our previous results imply that, at least for part of the high-redshift galaxy
population, the functional form of the attenuation curve differs significantly
from a Calzetti law. This is demonstrated by Fig.~\ref{fig_compmean}, which 
shows the best-fit Calzetti law for the ratio of the composite spectrum for 
$\gamma_{34} > -2$ to that for $\gamma_{34} < -2$. Neglecting possible 
differences in the age distribution of the stellar populations of both 
composite spectra, the resulting spectrum in the lower panel of 
Fig.~\ref{fig_compmean} should reflect only a difference in reddening, if the 
Calzetti law holds universally. As expected, the best-fit Calzetti law with 
$E_{B-V} = 0.08$ fails to reproduce the UV bump region. However, the
reliable wavelength range of the composites (from 1450 to 2900\,\AA{}) is too 
narrow to allow a significant deviation in the slope of the attenuation law 
to be detected. Moreover, this simple test does not provide any constraints 
on the reliability of the wavelength-dependent amount of attenuation 
predicted by the Calzetti law. Investigating the shape of the attenuation 
curve across a wide wavelength range would require a thorough study of a much 
larger data set and probably more sophisticated modelling. This is beyond the 
scope of the present work and is deferred to a future paper.

For now, we propose to modify the Calzetti et al. (\cite{CAL00}) law 
$k(\lambda)$ by adding a Lorentzian-like UV bump 
$c_3 D_{x_0,\gamma}(\lambda)$ (see Eq.~\ref{eq_drude} in Sect.~\ref{lorentz}) 
and derive the wavelength-dependent attenuation in magnitudes by 
\begin{equation}\label{eq_attcurve}
A(\lambda) = E_{B-V}\, (k(\lambda) + c_3 D_{x_0,\gamma}(\lambda)).       
\end{equation}
Since we have already used the Calzetti law as a baseline for the derivation 
of the FM-like parameters (see Table~\ref{tab_fmpar}), this approach allows 
us to fit the composite spectra of our subsamples. The Calzetti law appears
to work quite well as an average attenuation law for large samples of
high-redshift galaxies. Luminosity and SFR estimates derived by assuming the 
Calzetti law show good correspondence to values obtained by independent 
methods (e.g., Reddy \& Steidel \cite{RED04}; Erb et al. \cite{ERB06b}; Daddi 
et al. \cite{DAD07b}). Therefore, we assume that our modified Calzetti law 
describes, to zeroth order, on average dust attenuation in high-redshift 
star-forming galaxies with weak to moderate UV bumps. Characteristic values 
of $x_0$, $\gamma$, and $c_3$ are given in Table~\ref{tab_fmpar}.  

On the other hand, one may expect that the a priori use of a pure Calzetti
law for fitting SEDs of any high-redshift galaxy introduces systematic 
errors whenever the extinction curve produced by dust in a galaxy exhibits a 
significant UV bump. As a test, we constructed synthetic SEDs that are 
consistent with our data using Maraston (\cite{MARA05}) models reddened by a 
modified Calzetti law, according to a case `gm\_sb' UV bump (see 
Table~\ref{tab_fmpar}). The corresponding broad-band filter fluxes for 
different redshifts were then fitted by using the correct attenuation curve 
and a {\em pure} Calzetti law. The latter yielded about 5\% higher masses and 
about 10\% lower SFRs. For extreme but realistic cases, these discrepancies 
could be twice as large. The exact values depend on star-formation history, 
redshift (because of the differing filter coverage of the UV bump), and, of 
course, dust obscuration. The change in the rest-frame UV colours caused by 
the presence of a UV bump appears to be interpreted by a fitting routine that 
uses a standard Calzetti law as a signature of the older age of the stellar 
populations and/or a reduced dust reddening. For a typical high-redshift 
galaxy, the resulting deviations are almost negligible in view of the 
statistical and systematic uncertainties related to SED-fitting in general. 
In the following sections, we therefore discuss fitting results with the 
{\em Hyperz} code based on a standard Calzetti attenuation law to be 
consistent with other studies. Nevertheless, one should keep in mind that, 
for objects with the strongest 2175\,\AA{} absorption features in their 
spectra, there could be noteworthy systematic deviations in the estimated 
galaxy properties.

\subsection{Extinction curve and global properties of galaxies}
\label{galaxyprop}

\subsubsection{Star-formation rates and stellar masses}\label{sfrmass}

\begin{figure}
\centering 
\includegraphics[width=8.8cm,clip=true]{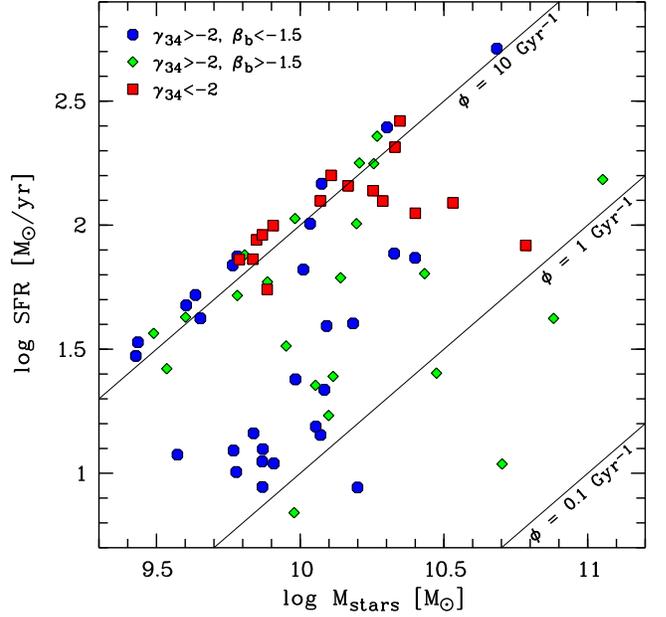}
\caption[]{Star-formation rate versus total stellar mass for the GMASS sample
galaxies with $1.5 < z < 2.5$. The three $\gamma_{34}$ and $\beta_{\rm b}$ 
selected subsamples are marked by different symbols (see legend). Specific 
SFRs of 10, 1, and $0.1$\,Gyr$^{-1}$ are indicated by lines. The upper left
corner of the diagram is not populated by galaxies because of the lower age 
limit of $100$\,Myr set for models with exponentially-declining SFRs
for the SED fitting.}
\label{fig_SFR_mass}
\end{figure}

Table~\ref{tab_avpar} lists average values of several global galaxy 
properties for the three GMASS subsamples defined in Sect.~\ref{frequency}.
In general, the monochromatic luminosities at 1500 and 2500\,\AA{} (rest 
frame) of GMASS galaxies with a robustly identified 2175\,\AA{} absorption
feature in their spectra do not differ from the average ones of the entire
GMASS sample. However, some differences exist in terms of the specific 
star-formation rate $\phi$, which is the SFR per unit total stellar mass 
$M_{\star}$ (derived from {\em Hyperz}, see Sect.~\ref{gmass}), as shown in 
Fig.~\ref{fig_SFR_mass}. In fact, GMASS SUBGs have 
$\langle M_{\star} \rangle = 1.4 \times 10^{10}\,\mathrm{M}_{\odot}$ and
$\langle \mathrm{SFR} \rangle \sim 100\,\mathrm{M}_{\odot}\,\mathrm{yr}^{-1}$,
so that their values of $\phi$ cluster around $7.1\,\mathrm{Gyr}^{-1}$. This 
is about twice the average $\phi$ of GMASS galaxies for which the signature 
of a UV bump in their spectra is weak or absent
($\langle \phi \rangle = 3.3\,\mathrm{Gyr}^{-1}$), whatever the observed UV 
slope. Interestingly, GMASS galaxies at $1.5 < z < 2.5$ with 
$\gamma_{34} > -2$ and blue UV continua tend to be less massive systems with 
lower SFRs than corresponding non-SUBGs with red UV continua.

A trend towards higher values of $\phi$ for SUBGs was also found by 
\cite{NOL07}. We note that the GMASS galaxies at $1 < z < 2.5$ considered 
here are a factor of two less massive than the FDF galaxies at $2 < z < 2.5$ 
in the \cite{NOL07} sample, but almost as massive as the FDF/K20 galaxies at 
$1 < z < 1.5$ in the \cite{NOL07} sample. Consistently, for the entire 
$4.5$\,$\mu$m-selected GMASS sample, which also comprises galaxies with 
photometry only and spectroscopic redshifts from other projects in GOODS-S, 
the average stellar mass drops from $\sim 2.4$ to 
$\sim 1.0 \times 10^{10}\,\mathrm{M}_\odot$ moving from $2 < z < 2.5$ to 
$1 < z < 1.5$, if $\mathrm{SFR} \ge 10$\,M$_\odot$/yr (lower limit for our 
spectroscopic sample). Apart from different sample selection criteria and 
different methods of estimating stellar masses\footnote{Comparison of stellar 
masses for galaxies in both samples suggests that the values used by 
\cite{NOL07} are systematically higher by about $0.1$\,dex, which is probably 
in general due to the different assumptions about the SFH (see also Drory et 
al. \cite{DRO05}).}, the average sample masses probably decrease towards lower 
redshifts due to the increasing fraction of low-mass objects in flux-limited 
samples and a decrease in the number density of massive star-forming galaxies 
related to the ``downsizing effect'' (Cowie et al. \cite{COW96}; Gavazzi \& 
Scodeggio \cite{GAV96A}; Gavazzi et al. \cite{GAV96B}).

\subsubsection{Morphology}\label{morph}

\begin{figure}
\centering 
\includegraphics[width=8.8cm,clip=true]{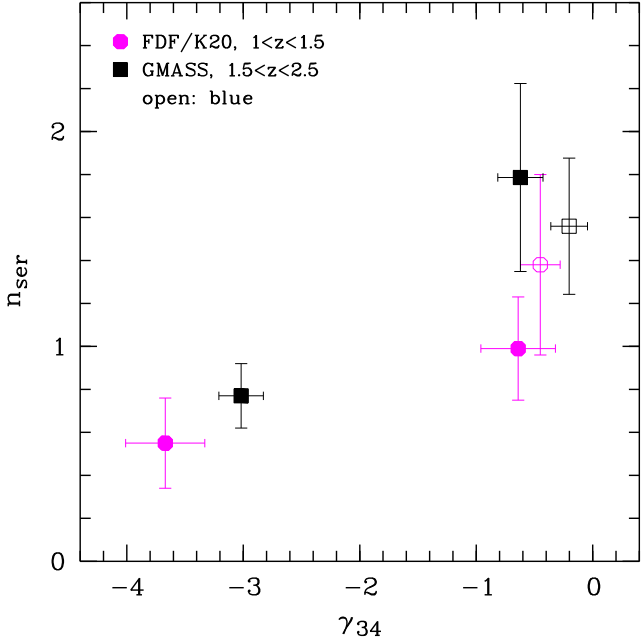}
\caption[]{Sersic index $n_{\rm ser}$ versus $\gamma_{34}$ for the three 
GMASS subsamples at $1.5 < z < 2.5$ and three subsamples of FDF and K20 
galaxies at $1 < z < 1.5$ discussed by \cite{NOL07}. The link between 
subsamples and symbols is illustrated in the legend. Open symbols indicate 
subsamples of galaxies with blue UV continua and weak 2175\,\AA{} features. 
Errors represent sample mean errors.}
\label{fig_av_nser_gamma34}
\end{figure}

Quantitative morphological studies of galaxies at $z \sim 2$ can be 
significantly affected by the relatively low surface brightness of, in 
particular, their outer regions, which can lead to erroneous 
morphological classifications. However, extremely deep ACS exposures are 
available for GOODS-South (see Sect.~\ref{gmass}), which allow us to analyse 
the rest-frame UV/U-band surface brightness distribution of our sample 
galaxies at least in an approximate way. Table~\ref{tab_avpar} shows the 
results for the effective radius $r_\mathrm{e}$, S\'ersic index 
$n_\mathrm{ser}$, elongation $b/a$, and the CAS parameters concentration $C$ 
and asymmetry $A$ (see Cassata et al. \cite{CAS05} for the definitions used) 
for our three GMASS samples at $1.5 < z < 2.5$. 

In general, the derived morphological parameters do not depend significantly
on the UV continuum properties. A remarkable exception is the average S\'ersic 
index for GMASS SUBGs $\langle n_\mathrm{ser} \rangle = 0.77 \pm 0.15$, 
which is conspicuously lower than the values $1.56 \pm 0.32$ and 
$1.79 \pm 0.44$ for coeval non-SUBGs with blue and red UV continua, 
respectively. Consequently, SUBGs appear to show radial distributions of the 
young stellar populations that are on average shallower than an exponential 
profile. On the other hand, half of the SUBGs have the minimum value allowed 
for the fit $n_\mathrm{ser} = 0.5$, which challenges the reliability of the 
fits. Moreover, there is no significant difference between our three GMASS 
subsamples in terms of the concentration $C$, which also depends on the radial 
surface brightness profile. However, this dependence is different, since $C$ 
is related to the ratio of the apertures containing 80\% and 20\% of the 
total flux, while $n_\mathrm{ser}$ describes the functional form of the 
radial light distribution. The concentration $C$ is obviously less sensitive
to the substructure of the radial profile than $n_\mathrm{ser}$, which 
appears to indicate irregular, very patchy rest-frame UV surface brightness 
profiles for low S\'ersic indices. A good argument for the reliability of the 
observed $\gamma_{34}$-related trend in $n_\mathrm{ser}$ is the result that a 
similar relation is present in the intermediate-redshift \cite{NOL07} sample 
at $1 < z < 1.5$ (see Fig.~\ref{fig_av_nser_gamma34}). Even if the physical 
interpretation of the $n_\mathrm{ser}$ values is difficult, the significance 
of the observed trend could in principle be used to search (in combination 
with other tracers) for SUBG candidates without available deep UV 
spectroscopy.   
     
The use of the CAS parameters for galaxy classification (see Conselice et al. 
\cite{CON03}) shows that the GMASS galaxies at $1.5 < z < 2.5$ appear to be 
a fair mixture of mergers/irregulars and disc galaxies, whatever 
$\gamma_{34}$. This distribution differs significantly from the one at 
$1 < z < 1.5$, where most of the SUBGs appear to have late-type morphologies 
(see \cite{NOL07}). Since mergers are frequent in the \cite{NOL07} sample of 
SUBGs at $2 < z < 2.5$, this suggests that the distribution of the young 
stellar populations in galaxies with LMC-like dust becomes smoother as time 
goes by.

\subsubsection{Narrow spectroscopic features}\label{spectrallines}

\begin{table}
\caption[]{Rest-frame equivalent widths in \AA{} of prominent lines in the 
composites of the three GMASS subsamples at $1.5 < z < 2.5$}
\label{tab_ews}
\centering
\begin{tabular}{l @{} c c c c}
\hline\hline
\noalign{\smallskip}
Line & Type$^\mathrm{a}$ &
$\gamma_{34} > -2$     & $\gamma_{34} > -2$     & $\gamma_{34} < -2$ \\
     &          &
$\beta_{\rm b} < -1.5$ & $\beta_{\rm b} > -1.5$ &                    \\ 
\noalign{\smallskip}
\hline
\noalign{\smallskip}
Si\,IV\,$\lambda$\,1400  & is & 
$+3.18 \pm 0.50$ & $+2.90 \pm 0.63$ & $+4.06 \pm 0.75$ \\
S\,V\,$\lambda$\,1502    & s  & 
$+0.55 \pm 0.09$ & $+0.60 \pm 0.07$ & $+0.83 \pm 0.11$ \\
Si\,II\,$\lambda$\,1526  & i  & 
$+1.97 \pm 0.12$ & $+1.95 \pm 0.17$ & $+2.73 \pm 0.28$ \\
C\,IV\,$\lambda$\,1550   & si & 
$+4.48 \pm 0.36$ & $+4.52 \pm 0.54$ & $+5.65 \pm 0.32$ \\
Fe\,II\,$\lambda$\,1608  & i  & 
$+1.36 \pm 0.15$ & $+1.72 \pm 0.12$ & $+1.72 \pm 0.27$ \\
He\,II\,$\lambda$\,1640  & sn & 
$-1.14 \pm 0.27$ & $-1.23 \pm 0.35$ & $-1.29 \pm 0.28$ \\
Al\,II\,$\lambda$\,1670  & i  & 
$+1.43 \pm 0.08$ & $+1.53 \pm 0.14$ & $+2.04 \pm 0.18$ \\
Al\,III\,$\lambda$\,1860 & i  & 
$+2.61 \pm 0.20$ & $+3.11 \pm 0.22$ & $+3.66 \pm 0.20$ \\
C\,III]\,$\lambda$\,1907 & n  & 
$-1.05 \pm 0.14$ & $-1.48 \pm 0.14$ & $-1.16 \pm 0.22$ \\
C\,II]\,$\lambda$\,2326  & n  & 
$-1.03 \pm 0.22$ & $-1.63 \pm 0.21$ & $-0.92 \pm 0.21$ \\
Fe\,II\,$\lambda$\,2344  & i  & 
$+1.97 \pm 0.35$ & $+2.07 \pm 0.19$ & $+2.95 \pm 0.26$ \\
Fe\,II\,$\lambda$\,2380  & i  & 
$+3.16 \pm 0.32$ & $+3.67 \pm 0.24$ & $+4.83 \pm 0.29$ \\
Fe\,II\,$\lambda$\,2600  & i  & 
$+4.79 \pm 0.48$ & $+5.15 \pm 0.62$ & $+5.85 \pm 0.68$ \\
Mg\,II\,$\lambda$\,2800  & is & 
$+4.01 \pm 0.75$ & $+4.02 \pm 0.73$ & $+6.59 \pm 0.76$ \\
Mg\,I\,$\lambda$\,2852   & s  & 
$+0.30 \pm 0.19$ & $+0.83 \pm 0.25$ & $+1.30 \pm 0.34$ \\
\noalign{\smallskip}
\hline
\end{tabular}
\begin{list}{}{}
\item[$^\mathrm{a}$] Line origin: s = stellar (photosphere and/or wind), 
i = interstellar absorption, n = nebular emission
\end{list}
\end{table}

\begin{figure}
\centering 
\includegraphics[width=8.8cm,clip=true]{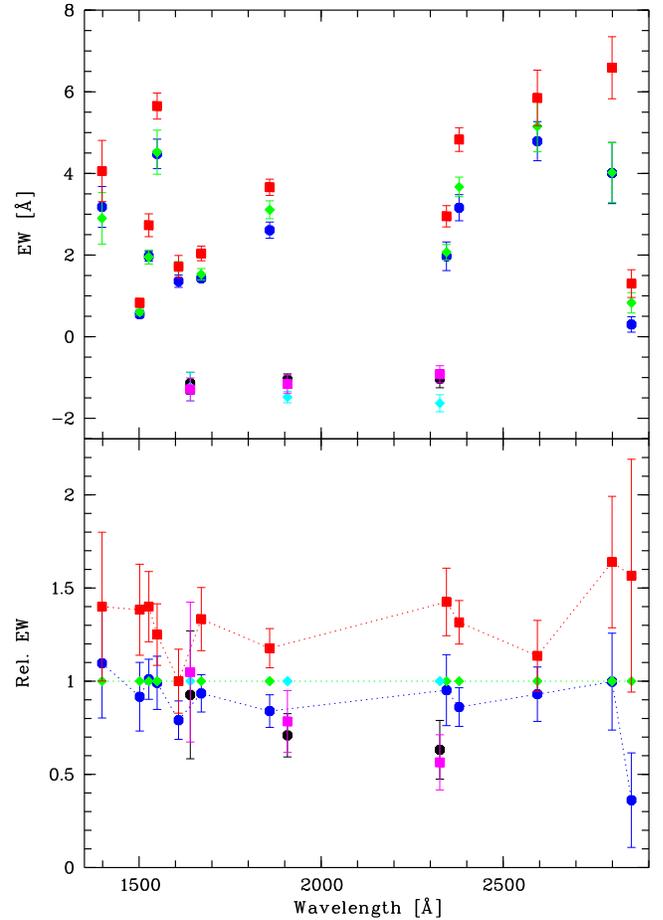}
\caption[]{Rest-frame equivalent widths of 15 spectral lines between 1350 
and 2900\,\AA{} for the composite spectra of three GMASS subsamples at 
$1.5 < z < 2.5$. Circles and lozenges mark blue and red galaxies with 
$\gamma_{34} > -2$. Squares indicate galaxies with $\gamma_{34} < -2$.
The $1\,\sigma$ error bars include the variance between the individual 
spectra, the errors of the continuum level definition, and the statistical 
uncertainties of the line strengths. {\it Upper panel:} Absolute equivalent 
widths in \AA{}. Emission lines have negative EWs. {\it Lower panel:} 
Equivalent widths relative to the values for the red subsample with weak 
2175\,\AA{} absorption features. Uncertainties on the reference values are 
considered by the error bars of the other samples. The relative EWs of the
absorption-dominated lines of each subsample are connected by dotted lines
in order to guide the eye.}
\label{fig_EWs}
\end{figure}

In the rest-frame UV domain mapped by the composite GMASS spectra, we can 
study 15 well-detected spectral lines from Si\,IV\,$\lambda$\,1400 to 
Mg\,I\,$\lambda$\,2852 (see Table~\ref{tab_ews}). The most prominent features 
are interstellar absorption lines and/or originate in stellar winds, and 
often show P~Cygni profiles. Equivalent widths (EWs) and their uncertainties 
are reproduced in Fig.~\ref{fig_EWs}, as well as relative EWs with respect to 
the GMASS subsample of galaxies at $1.5 < z < 2.5$ with $\gamma_{34} > -2$ 
and red UV continua. Comparison of the 12 absorption-dominated spectral lines 
of the GMASS blue and red subsamples of non-SUBGs reveals that the former 
type of galaxies has weaker features by only $11~(\pm 5)$\% on average. In 
contrast, GMASS SUBGs exhibit stronger features, on average, by 
$34~(\pm 5)$\%. On the other hand, the GMASS red subsample of galaxies with 
$\gamma_{34} > -2$ tend to have stronger emission-dominated lines than the 
others by 20--30\%. However, these emission lines have relatively uncertain 
EWs due to their weakness, even in composite spectra.  

Hence, we confirm that galaxies with secure detections of a broad absorption 
feature centred on 2175\,\AA{} in their spectra exhibit stronger 
low-ionisation interstellar absorption lines than those with a probable 
SMC-like extinction curve (\cite{NOL05}). This was interpreted as being due 
to a higher covering fraction of dusty neutral clouds in the former galaxies, 
which enables the carriers of the UV bump to survive. In this case, the 
low-ionisation interstellar lines trace the distribution of neutral gas 
clouds along the observer's sightline towards the hot stars producing the 
illuminating UV continuum radiation (Shapley et al. \cite{SHA03}). An 
alternative explanation for the presence of stronger interstellar
absorption lines is a broader velocity distribution (see Heckman et al. 
\cite{HEC98}). 

In addition, Fig.~\ref{fig_EWs} reveals that, in galaxies with a signature of 
a UV bump in their spectra, stronger absorption is inferred from all types 
of absorption lines, regardless of the wavelength and ionisation level. 
Hence, the higher absorption efficiency must be explained also for line 
components originating in stellar photospheres and winds. An enhanced 
metallicity is a likely interpretation. Since the strengths of wind features 
depend on metallicity (e.g., Leitherer et al. \cite{LEI01}; Mehlert et al. 
\cite{MEH02}), a higher metallicity could explain (at least in part) the 
significant absorption in the C\,IV\,$\lambda$\,1550\,\AA{} feature of 
SUBGs (cf. \cite{NOL05}). At least for the red galaxies in our sample, this 
interpretation is supported by the corresponding presence of UV nebular 
emission lines of lower strength (see Heckman et al. \cite{HEC98}; Noll et 
al. \cite{NOL04}) in the spectra of the same $\gamma_{34} < -2$ galaxies. 

Unfortunately, a determination of metallicity for the composite spectrum of 
each GMASS subsample considered here is not possible. Halliday et al.
(\cite{HAL08}) were able to determine a stellar metallicity equal to
$\sim 0.3$\,Z$_{\odot}$, based on the weak stellar
Fe\,III\,$\lambda$\,1978\,\AA{} index (Rix et al. \cite{RIX04}),
for a composite spectrum combining 75 galaxy spectra from the entire
GMASS survey. This iron-abundance, stellar metallicity is lower than the 
oxygen-abundance, gas-phase metallicity determined for UV-selected 
star-forming galaxies at $z \sim 2$ (Erb et al. \cite{ERB06a}), and 
metallicities determined for hot stars (Mehlert et al. \cite{MEH02}, 
\cite{MEH06}) and the ISM (Savaglio et al. \cite{SAV04}) in galaxies at 
similar redshifts, which exhibit values $\ga 0.5$\,Z$_{\odot}$. Halliday et 
al. (\cite{HAL08}) concluded that a light-element overabundance was being 
established in the selected GMASS star-forming galaxies as they were forming 
at $z \sim 2$. The higher EW of the Mg\,I\,$\lambda$\,2852\,\AA{} line
measured in GMASS galaxies that probably host the carriers of the UV bump
could be explained by a higher fraction of intermediate-age stars (cf. 
\cite{NOL07}). Unfortunately, uncertainties prevent us from establishing any 
difference in either the measured Mg-to-Fe EW ratio or the properties of the 
stellar populations (from our models) among our GMASS subsamples for 
$1.5 < z < 2.5$. Nevertheless, we note that the stellar metallicities of our 
star-forming, intermediate-mass galaxies at $1.0 < z < 2.5$ are not far from 
those of LMC stars.

\subsection{Dust grain populations emitting at mid-IR wavelengths}
\label{UV_IR}

\begin{figure}
\centering 
\includegraphics[width=8.8cm,clip=true]{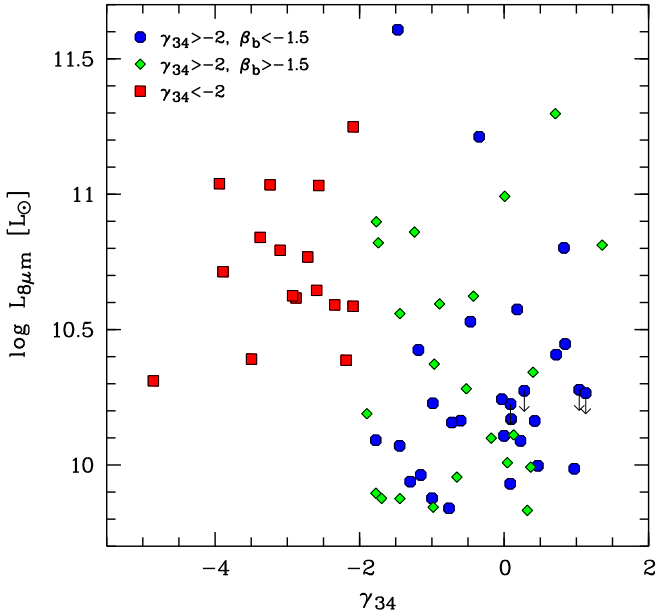}
\caption[]{Rest-frame luminosity at about 8\,$\mu$m $L_{8\,\mu{\rm m}}$ 
versus $\gamma_{34}$ for the three GMASS subsamples at $1.5 < z < 2.5$ (see 
legend for the symbols). Arrows indicate objects with 3\,$\sigma$ upper 
flux limits only.}
\label{fig_L8_gamma34}
\end{figure}

\begin{figure}
\centering 
\includegraphics[width=8.8cm,clip=true]{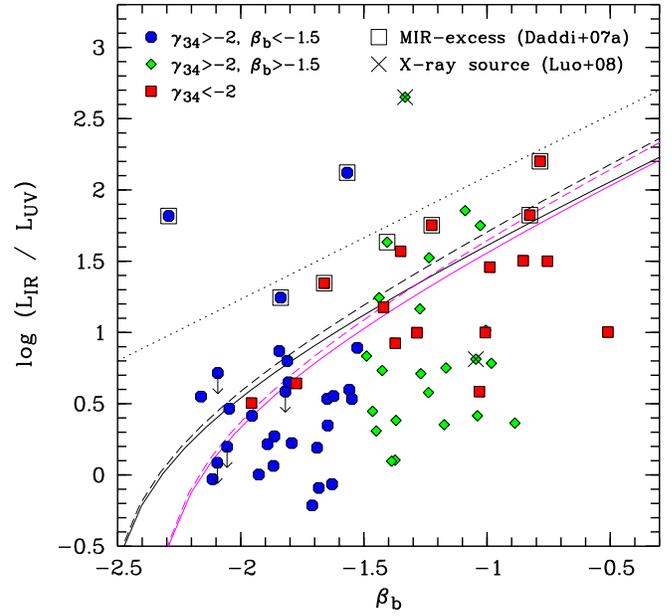}
\caption[]{Ratio of the IR luminosity $L_{\mathrm{IR}}$ and the UV luminosity
$L_{\mathrm{UV}}$ versus the UV-continuum reddening measure
$\beta_{\mathrm{b}}$ for the three GMASS subsamples at $1.5 < z < 2.5$
(see legend for the symbols). $L_{\mathrm{IR}}$ is derived from the 
24\,$\mu$m fluxes by using the Dale \& Helou (\cite{DAL02}) models, while 
$L_{\rm UV}$ is derived from SED-fitting in the wavelength range below
3000\,\AA{}. Galaxies that show particularly high luminosity ratios
in comparison to other galaxies with similar $\beta_{\mathrm{b}}$ are
separated by a dashed line. The mid-IR-excess objects of Daddi et al.
(\cite{DAD07a}) are indicated by big open squares. X-ray sources in the
catalogue of Luo et al. (\cite{LUO08}) are marked by big crosses. The X-ray 
source above the dotted line is not part of the Daddi et al. (\cite{DAD07a}) 
sample. Theoretical curves are given for solar-metallicity Maraston 
(\cite{MARA05}) models with either continuous SFR and age of 100\,Myr (black) 
or with $\tau = 3$\,Gyr $e$-folding time SFR and age of 3\,Gyr (magenta or 
grey). Solid lines are for a pure Calzetti et al. (\cite{CAL00}) law. Dashed 
lines show the results for a Calzetti law plus the `gm\_sb' UV bump in 
Table~\ref{tab_fmpar}.}
\label{fig_LirLuv_betab}
\end{figure}

The carriers of the UV bump probably consist of organic carbon and amorphous 
silicates (Bradley et al. 2005 and references therein), and seem to populate 
different environments of their host galaxy (see, e.g., Whittet \cite{WHI03}; 
Gordon et al. \cite{GOR03}; Fitzpatrick \cite{FIT04}; Clayton \cite{CLA04} 
and references therein). Another component of the dusty ISM, which is 
ubiquitous and also consists of carbonaceous material, is represented by the 
large polycyclic aromatic hydrocarbon (PAH) molecules (see Peeters et al. 
\cite{PEE04} and references therein). Far-UV pumped, vibrational emission of 
PAHs is generally held responsible for the well-known emission features at 
3.3, 6.2, 7.7, 8.6, and 11.2\,$\mu$m, which dominate the mid-IR spectra of 
many objects with associated dust and gas (L\'eger \& Puget \cite{LEG84}; 
Allamandola et al. \cite{ALL85}, \cite{ALL89}; Puget \& L\'eger \cite{PUG89}; 
Tielens et al. \cite{TIE99}, \cite{TIE00}). Unfortunately, there is no mid-IR 
spectroscopy for the GMASS sample, so that the physical conditions in the 
global environment where PAHs and the carriers of the UV bump exist cannot be 
inferred either quantitatively or qualitatively (from the comparison with the 
spectra of nearby objects). However, we can at least use the emission in the 
broad-band MIPS 24\,$\mu$m filter to estimate the alleged PAH emission around 
$\sim 8$\,$\mu$m (rest frame).

Observations in the MIPS 24\,$\mu$m filter are available for the 68 GMASS 
galaxies under study: fluxes range from 6 to 410\,$\mu$Jy, 14 galaxies being 
$2\,\sigma$ detections, and an additional four having only 3\,$\sigma$ upper 
limits. Interestingly, the four galaxies without detection exhibit 
$\gamma_{34} > -2$ and $\beta_\mathrm{b} < -1.5$. The MIPS 24\,$\mu$m fluxes 
are converted into rest-frame luminosities at $\sim 8$\,$\mu$m, where 
$L_{8\,\mu \mathrm{m}}$ is calculated from the filter-averaged $\nu f_{\nu}$ 
(cf. Daddi et al. \cite{DAD07a}). Relatively small cosmological 
$k$-corrections are obtained from the Dale \& Helou (\cite{DAL02}) IR 
templates, which are parameterised by the slope of the power-law distribution
of dust mass over heating intensity $\alpha$\footnote{Templates are chosen 
according to the mean relation between $\alpha$ (or $f_\nu^{60}/f_\nu^{100}$)
and the IR luminosity derived by Marcillac et al. (\cite{MARC06}) for local 
IR-luminous galaxies. We note that the Dale \& Helou (\cite{DAL02}) IR 
templates exhibit only small differences at 8\,$\mu$m (rest frame), so that 
small uncertainties are propagated to $L_{8\,\mu{\rm m}}$.}. Of the GMASS 
UV-luminous, intermediate-mass galaxies at $1.5 < z < 2.5$, those with a 
signature of the UV bump in their spectra appear to be more luminous (by 
$\sim 0.4$\,dex) at $\sim 8$\,$\mu$m (rest frame) than the other galaxies, 
even if only objects with similar UV-continuum reddening 
($\beta_\mathrm{b} > -1.5$) and stellar mass (cf. Table~\ref{tab_avpar}) are 
considered (see Fig.~\ref{fig_L8_gamma34}). The 8\,$\mu$m luminosities of the 
former galaxies are basically above $2 \times 10^{10}$\,L$_{\odot}$. 

For a clearer understanding of these results, we show the diagnostic diagram 
defined by the luminosity ratio 
$L_\mathrm{IR} / L_\mathrm{UV}$\footnote{$L_\mathrm{IR}$ is derived from the 
measured MIPS 24\,$\mu$m fluxes by the adoption of the Dale \& Helou 
(\cite{DAL02}) templates and the calibration of Marcillac et al. 
(\cite{MARC06}). $L_\mathrm{UV}$ is measured in the best-fit SED of 
{\em Hyperz} at wavelengths below 3000\,\AA{}.} and the UV-continuum slope 
$\beta_\mathrm{b}$ for the 68 GMASS galaxies at $1.5 < z < 2.5$ under study 
in Fig.~\ref{fig_LirLuv_betab}. Most of these galaxies are either located 
along the locus of nearby starbursts (cf. Meurer et al. \cite{MEU99}) or 
populate the region below this locus, where nearby ``normal'' star-forming 
galaxies lie (e.g., Kong et al. \cite{KON04}; Buat et al. \cite{BUA05}; 
Panuzzo et al. \cite{PAN07}). According to the interpretations given by the 
previous authors, the scatter of the GMASS SUBGs across the locus of nearby 
starbursts (the so-called ``Meurer law'', see Meurer et al. \cite{MEU99}) is 
consistent with the simple dust-screen model described in 
Sect.~\ref{frequency}. At the same time, the location of (red) galaxies with 
$\gamma_{34} > -2$ in Fig.~\ref{fig_LirLuv_betab} could indicate that the 
observed UV bumps are weakened due to more complex dust/star distributions 
(see \cite{NOL07} for a detailed discussion).

The a priori exclusion of optically identified AGN, still allows the presence
of continuum emission at mid-IR wavelengths caused by deeply obscured 
non-thermal sources. However, only two galaxies among the GMASS galaxies 
plotted in Fig.~\ref{fig_LirLuv_betab} have been detected in X-rays: none of 
them exhibits a significant UV bump. In fact, a non-detection in X-rays does 
not exclude a hidden AGN being present in our sample, as the X-ray stacking 
analysis of Daddi et al. (\cite{DAD07b}) indicates. Indeed, hidden AGN 
candidates identified as ``IR-excess objects'' (see Daddi et al. 
\cite{DAD07a}) are present among GMASS galaxies at $1.5 < z < 2.5$ with 
$\gamma_{34} < -2$ (25\%) or with $\gamma_{34} > -2$ (10\%)\footnote{One of 
the latter objects is actually part of the GOODS-S X-ray catalogue of Luo et 
al. (\cite{LUO08}) (see Fig.~\ref{fig_LirLuv_betab}).}. However, the moderate
$L_\mathrm{IR} / L_\mathrm{UV}$ for most of the GMASS galaxies in 
Fig.~\ref{fig_LirLuv_betab} suggest that the measured MIPS 24\,$\mu$m fluxes
are mainly associated with PAH emission and star formation. Moreover, our 
three GMASS subsamples do not show suspicious values and significant 
differences in IRAC colours that can discriminate between galaxies dominated 
by star formation and by AGN in the rest-frame near-IR (see Lacy et al. 
\cite{LAC04}; Stern et al. \cite{STE05}; Pope et al. \cite{POP08}). For 
example, SUBGs and non-SUBGs show on average $S_{8.0}/S_{4.5} = 0.8$ and no 
SUBG was found to be located above the Pope et al. (\cite{POP08}) AGN 
criterion of $2.0$. Finally, the Teplitz et al. (\cite{TEP07}) {\em Spitzer} 
IRS spectrum (source~2 in their Fig.~5) of the galaxy in the \cite{NOL07} 
sample with the strongest 2175\,\AA{} feature, a K20 galaxy at $z = 1.097$, 
indicates a typical, star-formation-dominated, luminous infrared galaxy. 
Unfortunately, this convincing check cannot be performed for other sample 
galaxies because of the lack of published mid-IR spectroscopy.
 
Additional support against an AGN-related origin of the observed mid-IR 
emission comes from the stacking analysis of MIPS fluxes in the 70\,$\mu$m 
filter for the three GMASS subsamples (D. Elbaz \& B. Magnelli, priv.comm.).
As a result, only $3\,\sigma$ upper limits between 600 and 800\,$\mu$Jy
are obtained. This excludes the suitability of Dale \& Helou (\cite{DAL02}) 
IR SEDs with power-law slopes of the dust mass distribution over heating 
intensity $\alpha < 1.2$, which peak at wavelengths $< 63$\,$\mu$m in 
$f_\nu$ and are more extreme than SEDs of typical star-forming galaxies 
(Dale \& Helou \cite{DAL02}). The typical values of $\alpha$ that we 
determine from the MIPS $24$\,$\mu$m fluxes and the calibration of Marcillac 
et al. (\cite{MARC06}) are about $1.7 - 1.8$.

We can therefore safely conclude that the bulk of the observed MIPS 
$24$\,$\mu$m flux of the GMASS galaxies under study comes from PAH emission, 
this emission being higher for objects with relatively strong 2175\,\AA{}
absorption features in their spectra. This enhanced PAH emission is probably
due to at least one of the following reasons:
1) a higher abundance of PAHs or a larger amount of dust;
2) a more efficient attenuation by dust as a result of the dust/star 
distribution (see \cite{NOL07} for a discussion);
3) a stronger illuminating radiation field, as a consequence of a larger UV 
throughput or a reduced dilution (i.e., increased closeness of dust to 
illuminating stars).
All of these independent interpretations are supported by our data to some 
level. In particular, the relation between the 8\,$\mu$m emission and the SFR 
or $\phi$ (see Table~\ref{tab_avpar}) suggests a link between star-formation
intensity and the amount of dust in the galaxy (cf. Wang \& Heckman 
\cite{WAN96}; Vijh et al. \cite{VIJ03}).

\section{Discussion}\label{discussion}

We have investigated spectral evidence of the broad absorption feature 
centred on 2175\,\AA{}, following the method developed by \cite{NOL05}. The 
\cite{NOL07} main sample of 88 actively star-forming, intermediate-mass
(i.e., with average stellar masses of $10^{10}-10^{11}\,\mathrm{M}_{\sun}$) 
galaxies at $1 < z < 2.5$ with appropriate spectra was enlarged by 
considering 68 GMASS galaxies at $1.5 < z < 2.5$. The gain consists not only 
in the larger size and the more homogeneous redshift coverage of the new 
sample. The additional GMASS galaxies have particularly high S/N optical 
spectra that bracket the redshifted wavelength region across the UV bump, so 
that a Lorentzian-like fitting of this broad feature can be performed. This 
allows us to study the shape of the UV bump and the link between its spectral 
signature (i.e., the parameter $\gamma_{34}$ introduced by \cite{NOL05}) and 
the properties (e.g., central wavelength and FWHM) of the fitted UV bumps.
Furthermore, the GMASS sample is large enough to enable subsampling: in 
composite spectra of sufficiently high S/N, EWs can be measured more robustly
for many narrower spectral features than in previous studies. These
absorption/emission lines originate in very different local environments,
such as stellar photospheres, wind regions, nebular regions, and the ISM.
Finally, available MIPS 24\,$\mu$m observations allow us to estimate the 
total amount of emission by dust at IR wavelengths and constrain the opacity 
characterising different GMASS subsamples. Hereafter, we discuss our main 
results in terms of galaxy structure, and the evolution of stellar
populations and the ISM.

\subsection{Differing extinction curves in high-redshift galaxies}
\label{propext}

We have produced robust evidence that at least 30\% of the UV-luminous,
intermediate-mass galaxies at $1 < z < 2.5$ contain dust that produces
an extinction curve with a broad absorption excess centred on 2175\,\AA{}
that differs from the standard SMC curve at UV wavelengths\footnote{Sample
selection criteria and survey sensitivity can hinder this fact, as suggested
by \cite{NOL07}.}. We refer to these objects, identified by 
$\gamma_{34} < -2$, as {\em s}ignificant {\em U}V {\em b}ump {\em g}alaxies
(SUBGs). With the exception of our previous investigations (\cite{NOL05}; 
\cite{NOL07}), the presence of the carriers of the UV bump in statistical 
samples of galaxies has been limited so far to intervening Ca\,II and 
possibly Mg\,II absorbers at intermediate/high redshifts (Malhotra 
\cite{MAL97}; Wang et al. \cite{WAN04}; Wild \& Hewett \cite{WIL05}). Our 
results imply that a non-negligible fraction of the observed actively 
star-forming galaxies at $1 < z < 2.5$ already has a UV extinction curve that 
differs from that characterising LBGs at similar and higher redshifts (cf. 
Vijh et al. \cite{VIJ03}). UV extinction curves in-between the average SMC one 
(with no UV bump but a steep far-UV slope) and the average LMC one (with a 
moderate UV bump and a less steep far-UV slope) appear to be common, whereas 
the average MW extinction curve (with a prominent UV bump and a less steep 
far-UV slope) does not. In particular, the shape of the UV bump inferred from 
the composite spectrum of SUBGs in the GMASS sample resembles that measured 
in the ``supergiant'' shell of ionised filaments LMC\,2, close to 
30\,Dor\footnote{This supergiant shell probably results from the combined 
action of stellar winds and supernova explosions over a timescale of
$10^7$\,yr (Caulet et al. \cite{CAU82}) and a unique magnetic field geometry,
consisting of coherent fields on spatial scales of $42 \times 24$\,pc
to $104 \times 83$\,pc with inferred projected field strengths
of $\sim 14 - 30$\,$\mu$G (Wisniewski et al. \cite{WIS07}).}.

Both the fraction of actively star-forming galaxies at $1 < z < 2.5$ with 
alleged UV bumps in their spectra, and the type of extinction curve (e.g., 
LMC versus MW average ones), depend on the dust/star distribution, which is 
generally unconstrained by existing observations of high-redshift galaxies 
(\cite{NOL07}). Extinction curves with a moderate UV bump are favoured if 
dust is mainly distributed above stars, as suggested for SUBGs, which are 
indicative of high covering fractions of young massive stars due to large EWs 
of strong interstellar absorption lines (Fig.~\ref{fig_EWs}). Moreover, the 
distribution of GMASS SUBGs in the 
$\beta_\mathrm{b}$--$L_\mathrm{IR} / L_\mathrm{UV}$ plane
(Fig.~\ref{fig_LirLuv_betab}) is aligned mainly along the opacity-dependent 
Meurer curve (Meurer et al. \cite{MEU99}) for starburst models with 
screen-like dust attenuation. Dust screens are expected if winds on a 
galactic scale or more localised superwinds are present (see, e.g., Murray et 
al. \cite{MUR05}). Galactic outflows are ubiquitous, particularly at high 
redshift (e.g., Weiner et al. \cite{WEI09}; see Heckman \cite{HEC05} and 
Veilleux et al. \cite{VEI05} for reviews). The presence of galactic winds or 
localised superwinds in GMASS galaxies, for instance, is supported by the 
high values of the specific SFR (Fig.~\ref{fig_SFR_mass}). Part of the 
expelled dust grains may settle in a potential minimum within the galaxy 
halo, as a result of simple dynamical arguments about the motion of 
interstellar grains out of the galactic disc (Greenberg et al. \cite{GREE87}; 
Barsella et al. \cite{BAR89}; Ferrara et al. \cite{FER90}, \cite{FER91}; 
Davies et al. \cite{DAV98}). The presence of poloidal magnetic fields may 
help to confine expelled dust grains (Beck et al. \cite{BEC94}), and, thus, 
preserve the SHELL-like configuration for the dust/star distribution that can 
describe dust attenuation in nearby and high-redshift starbursts (see Witt \& 
Gordon \cite{WIT00}; Vijh et al. \cite{VIJ03}). Increasing observational 
evidence points to the existence of organised magnetic fields in and around 
galaxies without AGN activity at intermediate/high redshifts, which are at 
least of comparable strength to those observed in nearby normal galaxies 
(Bernet et al. \cite{BER08}; Wolfe et al. \cite{WOL08}).

A higher fraction of objects can host the carriers of the UV bump, if dust is 
distributed mainly in the plane of the galaxy (as, e.g., in the radiative 
transfer models presented in Silva et al. \cite{SIL98} and Pierini et al. 
\cite{PIE04B}). As shown by \cite{NOL07}, a disc system with a MW extinction 
curve then exhibits $-3 < \gamma_{34} < 1$ only. The fraction of dust above 
the mid-plane of a galaxy can decrease as the SFR per unit area of the galaxy 
decreases (cf. Heckman \cite{HEC05}). Disc-dominated dust distributions 
become more likely as the cosmic SFR density decreases, i.e., towards 
redshifts progressively lower than unity (Madau et al. \cite{MAD96}). On the 
other hand, extinction curves more similar to those of the LMC or MW are also 
possible, if the dust/star distribution is patchy on very large scales (see 
\cite{NOL07}) or depends on the age of the individual stellar populations
(see Panuzzo et al. \cite{PAN07}). These dust/star distributions appear to 
explain the position of ``normal'' star-forming galaxies distinctly below the 
Meurer curve (Meurer et al. \cite{MEU99}; Kong et al. \cite{KON04}; Buat et 
al. \cite{BUA05}), where many non-SUBGs in our sample are located (see 
Fig.~\ref{fig_LirLuv_betab}).

Finally, we note that the extinction properties inferred in this study and in 
our previous ones refer to a galaxy-averaged environment. It is well known 
from studies of the MW, LMC, and SMC that bulk dust properties vary on much 
smaller scales than a galaxy scale length (down to sub-pc). Hence, the 
carriers of the UV bump may exist in particular regions of galaxies that do 
not exhibit spectral evidence of the 2175\,\AA{} feature. This may explain 
why a typical LMC extinction curve has so far been inferred only for the 
sightline towards GRB\,070802 at $z = 2.45$ (El\'iasd\'ottir et al. 
\cite{ELI08}), whereas GRB-host galaxies are commonly assumed to host 
SMC-like dust. It is clear that the stardust yields and the processes of 
creation and transformation of dust species and sizes in the general ISM play 
a fundamental role in explaining the variety of extinction curves in 
different galaxy environments. This is discussed in the next section.

\subsection{The carriers of the UV bump and other small-size dust components
at high redshifts}\label{carriers}

Some forty years after the discovery of the UV bump, proposals about the 
chemical composition of its carriers remain largely conjectural. If the 
observed properties of the UV bump strongly suggest that its carriers are in 
the small particle limit (i.e., $a << \lambda$), the abundance argument limits 
the set of possible constituent elements to C, Mg, Si, and Fe. Graphitic 
carbon is by far the most widely discussed and accepted material among these 
candidates, since it is the only one to exhibit a broad excess in the 
absorption cross-section at UV wavelengths (see Papoular \& Papoular
\cite{PAP09} for a recent discussion). However, as for any material, its 
properties must explain the variations in width of the UV bump, whilst 
conserving its peak wavelength (Whittet \cite{WHI03})\footnote{The width of 
the UV bump possibly varies owing to different mechanisms, including particle 
clustering, ice-mantle growth, compositional inhomogeneities, porosity 
variations, and surface effects. Our data do not give us a handle on any of 
these processes; however, we expect less particle clustering and mantle 
growth in carriers that are above the mid-plane of the galaxy (as in SUBGs), 
since the two phenomena take place in regions like molecular clouds, 
protostellar discs, etc.}. The random, hydrogen-free assembly of microscopic
sp$^2$ carbon chips forming a macroscopically homogeneous and isotropic
solid, which was proposed as a model carrier of the UV bump (Papoular \&
Papoular \cite{PAP09}), is able to explain these properties taking into 
account astrophysical conditions in the ISM. The most promising, alternative 
carbon-based materials are synthetic nanoparticles condensed in a 
hydrogen-rich atmosphere and polycyclic aromatic hydrocarbons (PAHs), which 
exhibit a rise in the absorption cross-section across the region in which the 
UV bump originates. In addition, oxygen-rich materials have been proposed as 
carriers of the UV bump. In particular, silicate-based materials appear the 
most probable candidates, although this may require unreasonable fine-tuning 
of the size distribution (Whittet \cite{WHI03}). If graphitic grains (at 
least partly) are responsible for the UV bump, the main effect of the 
presence of PAHs (and probably silicates) will be then to increase the 
absorption cross section in particular at the wings of the UV bump, causing 
the feature to become less prominent. A variable mixture of the three 
components could explain variations in the height and width of the UV bump 
among observed extinction curves. Thus, it is safe to assume that the 
carriers of the UV bump are a mixture of amorphous silicates and carbonaceous 
material, as indeed observed in solar-system interplanetary particles 
(Bradley et al. \cite{BRA05})\footnote{Carbon-rich and silicate materials are 
made in different circumstellar environments (see text later on). How and 
where the two components, originating in different production sites, were 
combined within the same dust particles remains unclear. Hence it is 
uncertain whether the history of solar-system particles can be regarded as 
typical of those in remote galaxies. Moreover, Bradley et al. (\cite{BRA05}) 
could not distinguish whether the UV bump is due to organic carbon or 
amorphous silicates (or both), nor could exclude other possible UV-bump 
carriers.}.

According to the successful dust model of D\'esert et al. (\cite{DES90}),
the far-UV non-linear rise in the extinction curve can be ascribed
mostly to large molecules (radius $a \sim 0.4 - 1$\,nm), i.e., PAHs,
whereas carbonaceous nanoparticles ($a \sim 1 - 10$\,nm) called very small
grains (VSGs) are responsible for the sharp absorption excess at 2175\,\AA{},
and big grains (BGs, $a \sim 10 - 100$\,nm) cause most of the extinction
at UV wavelengths, but in a featureless way and with decreasing efficiency
from the near- to the far-UV domain. Furthermore, PAHs are the carriers of
the aromatic infrared bands between 3 and 17\,$\mu$m, whereas VSGs
are responsible for the continuum emission at mid-IR wavelengths. Differences
in the far-UV slope and/or the strength of the UV bump are related to zeroth 
order differences in the fraction of big grains and in the VSG-to-PAH 
abundance ratio of the dust. The latter arise from both dust production, 
either in stellar outflows (i.e., stardust) or in the general ISM, and dust 
processing in the general ISM (Whittet \cite{WHI03} and references therein). 
Regardless of whether the carriers of the UV bump originate in stars, the 
main unknown is the abundance required to alter in the appropriate way the 
shape of the average extinction curve of a galaxy.

Silicates (mostly amorphous) are produced in outflows of oxygen-rich cold
stars, such as O-rich asymptotic giant branch (AGB) stars and red giant and
supergiant stars (Whittet \cite{WHI03} and references therein). In contrast,
amorphous carbonaceous grains and PAHs are produced in outflows of
carbon-rich cold stars, i.e., AGB carbon stars\footnote{The amorphous carbon
grains ejected by C-rich stars need to be at least partly graphitised by
interstellar processes to produce the distinctive absorption feature
at 2175\,\AA{} (Whittet \cite{WHI03} and references therein). This would take
about 100\,Myr if such a process is driven by UV radiation (see Sorrell 
\cite{SOR90}; Mennella et al. \cite{MEN96}, \cite{MEN98}).}. In any event,
intermediate-mass stars ($1 < M < 8$\,M$_\odot$) lose most of their mass
during their AGB phase (e.g., Iben \& Renzini \cite{IBE83}). Hence, they are
major producers of dust grains, possibly including those responsible for
the UV bump. Besides the winds of AGB stars, the expanding envelopes of
Type II supernovae (SNe) represent an O-rich environment where silicate dust
can form (Whittet \cite{WHI03}). According to Zhukovska et al. 
(\cite{ZHU08}), SNe could also be important for the production of carbon 
dust, mostly in crystalline form (i.e., graphitic-like). In any case, the 
importance of SNe for the dust production in galaxies is still highly 
uncertain (Bianchi \& Schneider \cite{BIA07} and references therein). AGB 
stars and their outflows appear with a short delay after the beginning of star
formation, i.e., some $\sim 30$\,Myr, which is the lifetime of a 8\,M$_\odot$ 
star. AGB stars more massive than $\sim 3$\,M$_\odot$ never become carbon 
stars (e.g., Renzini \cite{REN08} and references therein), can hence 
potentially produce silicate stardust but not C-rich stardust. For the latter 
grains to appear, one needs to wait for stars less massive than 
$\sim 3$\,M$_\odot$ to reach the AGB, or $\sim 300$\,Myr after the beginning 
of star formation. Moreover, low metallicity favours the appearance of carbon 
stars, because less carbon needs to be dredged up in AGB stars to reach C/O 
ratios of above unity, and turn them into carbon stars (Renzini \& Voli 
\cite{REN81}). In summary, the theory of stellar evolution foresees that 
silicates are more promptly released than carbon-rich grains, and carbon 
grains are more abundantly produced in a low-metallicity environment.

With the persistent uncertainty in the chemical nature of the UV-bump 
carriers, their stellar producers, and their possible re-processing in the 
general ISM, much of the following discussion is speculative. It is safe to 
say that the EWs of several emission/absorption features of carbon and 
silicon imply that the chemical enrichment of these dust constituents has 
proceeded significantly in the SUBGs in our GMASS sample (see 
Table~\ref{tab_ews}). Additional evidence is provided by the emission at 
8\,$\mu$m (rest frame), if this emission is dominated by PAHs. In particular, 
emission features around 8\,$\mu$m (rest frame) are associated with C-rich 
stars (Whittet \cite{WHI03} and references therein). Thus, AGB outflows 
probably play a dominant role in the build-up of the carriers of the UV bump 
in massive galaxies at high redshift (\cite{NOL05}; \cite{NOL07}), as well as 
in the build-up of PAHs in local (Galliano \cite{GAL08}) and high-redshift 
galaxies alike. Consistently, dust exclusively formed in the ejecta of 
core-collapse SNe and exposed to sputtering by the hot gas appears to produce 
extinction curves that do not exhibit a UV bump (see Bianchi et al. 
\cite{BIA07}; Nozawa et al. \cite{NOZ08} and references therein).

The absence of the UV bump in the SMC challenges the idea that carbon grains 
represent its carriers, since the SMC is very rich in carbon stars and yet no 
UV bump is present there. The lack of a UV bump could also be caused by 
efficient grain destruction (Gordon et al. \cite{GOR03}; Sloan et al. 
\cite{SLO08}) favoured by lower dust column densities and harder radiation 
fields due to the low metallicity. If carbonaceous carriers of the UV bump 
are more sensitive to harsh environments than silicate ones, then the UV bump 
strength may be lower in carbon-dust dominated low-metallicity environments 
than in high-metallicity galaxies, where the dominant population of silicates 
may also act as carriers of the UV bump. The latter is possible for our 
sample of SUBGs (cf. Sect.~\ref{spectrallines}). On the other hand, these 
galaxies appear to contain a large amount of PAHs (see Sect.~\ref{UV_IR}), 
which are carbonaceous in nature. 

The destruction of refractory dust proceeds predominantly in the intermediate 
warm phase of the ISM (McKee \cite{McK89}; Jones et al. \cite{JON94}). This 
phase of the ISM has probably a large filling factor in actively star-forming 
galaxies at high redshift. In the low-density phases of the ISM, where SN 
explosions generally occur (at least in the local Universe), the relative 
abundance of PAHs with respect to VSGs is higher than in the high-density 
phases of the ISM. This was also found to be true for the diffuse, 
illuminated part of dense photodissociation regions (Compi\'egne et al. 
\cite{COM08}), whose chemistry is controlled by the external UV radiation 
field. Thus, we can assume that, in diffuse (dense) regions of the ISM of 
UV-luminous, massive galaxies at high redshift, narrow (broad) UV bumps are 
observed as a consequence of the higher (lower) fraction of PAHs. This trend 
is observed in the MW (Fitzpatrick \& Massa \cite{FIT86}; Cardelli \& Clayton 
\cite{CAR91}; Valencic et al. \cite{VAL04}), although it is usually explained 
in terms of ice mantle growth in the dense regions (Mathis \cite{MAT94}). 
Unfortunately, a similarly clear trend has not been observed in the LMC 
(Misselt et al. \cite{MIS99}). On the other hand, different dust components 
may have different spatial distributions on scales ranging from pc to kpc, 
and, thus, be exposed to different stresses caused by the UV radiation field 
and shocks. Unfortunately, the few existing computations of dust expulsion 
and erosion reach conflicting conclusions about the distances at which 
graphite and silicate grains are located and the equilibrium distances 
reached by grains of different sizes (cf. Bianchi \& Ferrara \cite{BIA05}; 
Aguirre et al. \cite{AGU01a}, \cite{AGU01b}). Our analysis suggests that 
carriers of the UV bump populate regions at some distance from the mid-plane 
of a galaxy at least in SUBGs. They might have survived expulsion or been 
produced by erosion during expulsion of larger carbonaceous grains. 

Whatever the carriers of the UV bump are, it is reasonable to expect that a
higher fraction of them are hosted by more evolved galaxies at high redshift.
If they are mostly silicates, galactic nucleosynthesis would be required to 
accumulate sufficient silicon and oxygen and enable them to condense onto 
grains in massive AGB star winds. If they are mostly carbonaceous, stars of
lower mass than $\sim 3$\,M$_{\sun}$ have to reach the AGB ($\sim 300$\,Myr) 
and start producing C grains, their production rate peaking when 
$\sim 2$\,M$_{\sun}$ stars reach the AGB, about 1\,Gyr later. If the carriers 
of the UV bump are not stardust, they are produced in the general ISM on 
timescales of the order of $0.5$\,Gyr (Whittet \cite{WHI03} and references 
therein). The dust production timescales can become significantly shorter, if 
SNe of Type II (which could occur at very high redshift; see Liang \& Li 
\cite{LIA08}; Nozawa et al. \cite{NOZ08}) or the slow winds of their 
precursors (red supergiants) play a role. In any case, the carriers of the UV 
bump have to spread and survive across and above the disc (possibly as high 
as 10 kpc), where part of them (or their parent grains) are transported to 
speeds of 10 to 500\,km\,s$^{-1}$ by galactic winds and superwinds. Finally, 
if the presence of the carriers of the UV bump is ruled not only by their 
accumulation (in stars or in the ISM) but also by their survival (e.g., 
favoured by self-shielding; see \cite{NOL05}), some threshold in either 
dust-to-gas ratio or metal abundance may have been reached (see Gordon et al. 
\cite{GOR03}).

\section{Conclusions}\label{conclusions}

We have selected a total of 78 UV-luminous galaxies at $1 < z < 2.5$ with 
average stellar mass of about $10^{10}\,\mathrm{M}_{\sun}$ from GMASS, a deep 
spectroscopic survey of IRAC $4.5$\,$\mu$m-flux limited galaxies at 
$z \sim 2$ in the GOODS-S (Kurk et al. \cite{KUR08A}). This sample extends to 
186 the number of actively star-forming, intermediate-mass (i.e., with 
stellar masses of $10^{10} - 10^{11}\,\mathrm{M}_{\sun}$) galaxies at 
$1 < z < 2.5$ with high S/N optical spectra and multiwavelength photometry, 
selected from the FDF, K20, and GDDS samples and investigated by 
\cite{NOL07}. Its heterogeneous nature assures us that the total sample 
probes the population of UV-luminous, massive galaxies at these redshifts 
comprehensively, beyond the limitations of the selection criteria of 
individual samples.

The properties of the UV extinction curve, i.e., far-UV slope and presence 
(or absence) of the broad absorption feature centred on 2175\,\AA{} 
(``UV bump''), were constrained by a parametric description of the spectra by
mapping the rest-frame UV spectral energy distributions of the individual 
galaxies. We have interpreted the data by applying a suite of different 
models combining radiative transfer and stellar population synthesis. For the 
68 GMASS spectra at $1.5 < z < 2.5$ in particular, additional Lorentzian-like 
``Drude'' fitting of the UV bump was performed. The overall analysis placed
constraints on the dust population of large molecules and small grains 
responsible for the extinction at UV wavelengths.

Hereafter, we summarise the main results of our analysis and place them in 
the context of our previous results.
\begin{itemize}
\item[--] About 30\% of the spectra of actively star-forming, 
intermediate-mass galaxies at $1 < z < 2.5$ undoubtedly exhibit broad 
absorption features centred on $\sim 2175$\,\AA{}, regardless of redshift. We 
term such galaxies SUBGs ({\em s}ignificant {\em U}V {\em b}ump 
{\em g}alaxies). Taking into account their measured UV slopes, we conclude 
that the extinction curves characterising these galaxies differ from those 
observed in the SMC, as well as from those characterising nearby starburst 
galaxies and typical LBGs at $2 < z < 4$.
\item[--] If attenuation at UV wavelengths is dominated by dust distributed 
above the stellar body of a galaxy, as suggested at least for SUBGs, 
effective UV extinction curves in-between those typically observed in the 
SMC and LMC characterise UV-luminous, intermediate-mass galaxies at 
$1 < z < 2.5$. This dust/star configuration originate from the action of 
galactic winds or localised superwinds. Stronger intrinsic UV bumps are 
possible for an enhanced mixing of dust and stars (probably in a disc), 
patchier dusty screens, age-dependent extinction effects, and significantly 
diverse regions in terms of star formation and dustiness within a galaxy. 
Since this cannot be excluded, especially for non-SUBGs (for which the ratio 
$L_\mathrm{IR} / L_\mathrm{UV}$ is significantly lower than expected for 
dust-screen dominated galaxies), the fraction of high-redshift galaxies with 
LMC-type extinction curves can be higher than inferred from the apparent 
bump strength.
\item[--] The UV bumps of the analysed UV-luminous high-redshift galaxies 
exhibit about 60\% of the width typically found in the LMC and Milky Way.
In the local Universe, similar widths are found only for some sightlines
towards the supergiant shell of ionised filaments LMC\,2 near 30\,Dor. This
suggests similar environments in terms of radiation field, dust-to-gas mass
ratio, and population of dust grains and molecules.
\item[--] The presence of multiple carriers of the UV bump (probably organic 
carbon and amorphous silicates) in a galaxy argues that AGB stars have
already started to return large amounts of dust to the ISM. The required
chemical enrichment is particularly evident for GMASS SUBGs at 
$1.5 < z < 2.5$, for which relatively large equivalent widths can be measured 
for several UV absorption features, mainly of interstellar origin, produced 
by Si, S, C, Al, Fe, and Mg.
\item[--] SUBGs are more luminous at 8\,$\mu$m (rest frame) than most of the 
other GMASS galaxies at $1.5 < z < 2.5$, which, on the other hand, have lower 
average star-formation rates. This is circumstantial evidence that 
carbonaceous dust has already been synthesised in relatively large amounts at 
these redshifts, since polycyclic aromatic hydrocarbons (which originate in 
AGB stars) probably dominate the emission at 8\,$\mu$m.
\item[--] Large amounts of UV radiation and shocks must be present in our 
sample galaxies. This is particularly true for SUBGs, since their specific 
star-formation rates are higher. Survival of the small-size dust component in 
such harsh environments is helped by the higher injection rates and more 
efficient dust self-shielding. Consistently, UV-luminous, massive galaxies at 
$1 < z < 2.5$ with LMC\,2-like dust suffer from higher attenuation at UV 
wavelengths and, according to our models, host larger amounts of dust.
\end{itemize}

The frequent occurrence of significant UV bumps in spectra of high-redshift
galaxies appears to imply that these galaxy SEDs could be reproduced more 
accurately by models with attenuation curves that differ from the Calzetti 
law, which does not account for the presence of a UV bump. The investigation 
of the attenuation curves of star-forming, high-redshift galaxies across a 
wider wavelength range is deferred to a future paper. However, the present 
study indicates that a Calzetti law plus a Lorentzian-like bump reproduces 
the spectra of SUBGs at least for the wavelength range $1450 - 2900$\,\AA{}. 
Accounting for the presence of a UV bump in this way involves systematic but 
small changes in crucial properties such as the SFR in SUBGs.

\begin{acknowledgements}
Many thanks go to David Elbaz and Benjamin Magnelli for the 70\,$\mu$m 
stacking using the GOODS-S data of David Frayer and collaborators.
DP thanks Adolf Witt for useful discussions on the carriers of the UV bump.
Finally, we thank the anonymous referee for her/his helpful comments that 
improved this paper. The spectra of the GMASS (ESO large programme 
173.A-0687), FDF, and K20 samples are based on observations obtained with 
FORS at the ESO VLT, Paranal, Chile. The latter spectra as well as the Hubble 
ACS images of the GOODS-S were retrieved from the ESO/ST-ECF Science Archive 
Facility. The GDDS spectra stem from observations with GMOS at the Gemini 
North Telescope, Mauna Kea, USA. SN is funded by the \emph{Agence Nationale de 
la Recherche} (ANR) of France in the framework of the D-SIGALE project. JDK 
is supported by the \emph{Deut\-sche For\-schungs\-ge\-mein\-schaft} (DFG), grant 
SFB-439.
\end{acknowledgements}

\end{document}